\documentclass[%
 aip,
 amsmath,amssymb,
preprint,%
 longbibliography,
 floatfix,
]{revtex4-1}

\usepackage{graphicx}
\usepackage{dcolumn}
\usepackage{bm}

\usepackage[utf8]{inputenc}
\usepackage[T1]{fontenc}
\usepackage{mathptmx}
\usepackage{etoolbox}
\usepackage{xcolor}
\usepackage{float}
\usepackage{threeparttable}

\makeatletter
\def\@email#1#2{%
 \endgroup
 \patchcmd{\titleblock@produce}
  {\frontmatter@RRAPformat}
  {\frontmatter@RRAPformat{\produce@RRAP{*#1\href{mailto:#2}{#2}}}\frontmatter@RRAPformat}
  {}{}
}%
\makeatother
\begin{document}

 
\title{A Large Eddy Simulation Study of Turbine Pitch Motion Effects on Power Generation in Fully 
Developed Floating Wind Turbine Arrays}
\author{Yishu Zhang}
\affiliation{ 
Department of Mechanical Engineering, Johns Hopkins University, Baltimore, Maryland 21218, USA
}%

\author{Dennice F. Gayme}%
\affiliation{ 
Department of Mechanical Engineering, Johns Hopkins University, Baltimore, Maryland 21218, USA
}%

\author{Charles Meneveau}
\thanks{\textbf{Author to whom correspondence should be addressed:} meneveau@jhu.edu}
\affiliation{ 
Department of Mechanical Engineering, Johns Hopkins University, Baltimore, Maryland 21218, USA
}%
\date{\today}

\begin{abstract}
Interactions between floating wind turbines and the marine atmospheric boundary layer affect turbine wakes and power output. This study examines the impact of turbine representations and platform pitching motion on power predictions for aligned floating wind turbine arrays. Large-eddy-simulations (LES) are performed for a fully developed wind turbine array boundary layer with periodic streamwise and spanwise boundary conditions and moving surface drag (MOSD) at the bottom-surface. A proposed power-curve-aware extension of the conventional actuator disk model (PCADM) is compared with the traditional constant coefficient actuator disk model (CCADM),  valid in principle only during Region 2 operation. Harmonic pitching motion matched to a representative ocean wave frequency is prescribed for IEA 15MW turbines mounted on the UMaine VolturnUS-S platform. Results demonstrate that the CCADM produces unphysical power overshoot when relative velocities induced by the harmonic pitching periodically exceed rated speed, while the PCADM eliminates these unrealistic excursions from the actual turbine operating range. Various wind–wave conditions and relative pitching motions for an aligned farm with PCADM turbines are analyzed, including in-phase, opposite phase and incommensurate phase row-to-row relative turbine motion. Results reveal that depending on operational regions, turbine pitching motion can either increase or decrease mean turbine power output. Among various relative phases, in-phase motions produces larger fluctuations at the farm-scale, due to stronger turbine–turbine correlations. These findings highlight the importance of accounting for turbine motion with associated varying operational regions when modeling turbine and wind farm power production in LES of offshore floating wind farms using actuator disk representations.
\end{abstract}

\maketitle



\section{\label{sec:Intro}Introduction}

Offshore floating wind turbines (OFWTs) operate in a highly complex and dynamic marine environment, where the coupled interactions among wind, waves, and turbine–platform dynamics play a central role in determining overall performance. In contrast to fixed-bottom wind turbines (FBWTs), OFWTs are not rigidly anchored to the seabed and can respond to environmental loads through platform motion. The motion of a floating platform typically has six degrees of freedom, comprising three translational motions, surge, sway and heave, and three rotational motions, roll, pitch and yaw. Among these motions, surge and pitch are generally considered the dominant or most influential platform responses for floating offshore wind turbines\cite{Subbulakshmi2022,zhou2025optimized,Vire2026FOWE}. The present study focuses specifically on modeling in the presence of turbine/platform pitching motion and its impact on mean and fluctuating power production of turbines and the wind farm (although the results will be argued to be equally applicable to surge motions).

Extensive efforts have been devoted to understanding wake dynamics \cite{Rockel2014, messmer2024enhanced,li2024resolvent} and power production \cite{Yang2014, Porchetta2019, Wen2018} in offshore wind systems using numerical simulations as well as experiments for both FBWTs and OFWTs. Using Large Eddy Simulation (LES) with actuator disk turbine models, Yang et al. \cite{Yang2014} reported that the average power output of 
an extended FB offshore wind farm in the presence of a downwind swell increases by $18.8\%$ and $13.6\%$ at moderate wind speeds of $7 \mathrm{m/s}$ and $10 \mathrm{m/s}$, respectively. For a single standalone stationary turbine, using actuator line model LES, AlSam et al\cite{AlSam2015} found that swell can enhance the turbine’s power output and extends the wake length. 
Using a free vortex model, Wen {\it et al.} \cite{Wen2017,Wen2018}  examined the influence of surge and pitch motions on OFWT performance across a broad range of tip-speed ratios, motion frequencies, and motion amplitudes. This study demonstrated that platform motion can either enhance or reduce mean power production depending on the tip-speed-ratio (TSR), although the time-averaged $C_P$ generally remained lower than that of the fixed-bottom case. 
All of these simulation-based studies assumed fixed thrust and power coefficients ($C_T$ and $C_P$, respectively).

Experimental studies by Sant {\it et al.} \cite{Sant2015} likewise reported both increases and decreases in time-averaged power production of OFWTs relative to equivalent fixed-bottom turbines, with the largest deviations occurring under high-frequency wave conditions. Their analysis further indicated that the instantaneous power varies primarily with the platform surge velocity through changes in the rotor-relative wind speed. More recently, Duan {\it et al.} \cite{duan2025floating} developed a theoretical framework based on the assumptions of a constant power coefficient and uniform inflow velocity across the rotor. Their analysis demonstrated that pitch motion can either increase or decrease the rotor-available power relative to a fixed-bottom turbine, depending on the Strouhal number. Recent developments in floating wind turbine modeling have increasingly incorporated detailed and realistic turbine control for individual turbines. For example, Tran {\it et al.}~\cite{Tran2026} presented and validated an aero-hydro-servo-elastic model of the IEA~15 MW reference turbine on the University of Maine VolturnUS-S platform~\cite{Allen_et_al_UMaine_VolturnUS_15MW}, incorporating generator-torque and blade-pitch control strategies for below- and above-rated operating conditions. 

In most prior laboratory experimental studies, the lack of wind turbine model feedback control does not enable  the system to track a turbine's real power curve. Instead, the rotational speed is typically governed indirectly through the electrical loading of the generator\cite{energies2017_10_908,wei2022phase,Rockel2014}, which in turn determines the TSR for a given inflow velocity. For example, Bastankhah and Port{\'e}-Agel\cite{energies2017_10_908} used electronically adjustable loading to systematically vary the rotor speed, allowing the power coefficient $C_p$ to be characterized as a function of TSR. Similarly, Wei and Dabiri\cite{wei2022phase} applied a comparable approach, but with manually adjusted electrical loads, producing a range of TSR spanning both optimal and off-design conditions. Given the complexities of specifying  turbine rotation control, many laboratory experimental studies\cite{Rockel2014,messmer2024enhanced,duan2025floating} have assumed constant, optimal thrust and power coefficients ($C_T$ and $C_p$) in which resistive loading was tuned to maintain a nominally constant TSR across different operating conditions. Across these approaches, no active control system of rotor-speed or aerodynamic-load in response to varying inflow conditions is employed. The electrical load primarily sets generator torque, which passively determines rotor speed. Consequently, the resulting $C_T$ and $C_P$ remain fixed at values corresponding to either optimal or off-optimal operating conditions, regardless of variations in the inflow. 

The assumption of fixed coefficients is also commonly adopted in laboratory experiments using porous disks \cite{Howland2016Wake, espana2011spatial} which implies assuming a constant $C_T$ value. Similarly,  LES studies of wind farms \cite{Calaf_Meneveau_Meyers_2010, Stevens_Meneveau_2017, Xiao_Yang_2019, Yang2014}, turbines are often modeled as actuator disks operating at constant, near optimal $C_T$ and $C_p$. While this assumption may be reasonable for fixed-bottom turbines operating at speeds below the rated wind speed, it becomes less appropriate for OFWTs, where wave-induced turbine motion can induce large time-periodic relative wind velocities experienced by the turbine that can lead to cross-overs between operating regions. To address this limitation of existing ADMs, the present study develops a power-curve-aware actuator disk model for LES framework, which explicitly incorporates transitions between turbine operating  regions and their dependence on time-varying inflow conditions, enabling a more realistic representation of turbine operation under dynamic OFWT environments.

The remainder of this paper is organized as follows. Sec.~\ref{sec:wind-turbine} reviews the traditional constant coefficient actuator disk model (CCADM) and describes the proposed power curve-aware extension, named PCADM. PCADM enables representation of the different turbine operational regions in LES using actuator disk models (ADMs). The relative motion between floating turbines and incoming wind is considered in Sec.~\ref{sec:relativemotion}. Sec.~\ref{sec:LES} describes the LES framework, simulation setup, and parameter choices used to examine predictions from the PCADM approach. Simulation results dealing with effects of PCADM on power output are provided in Sec.~\ref{sec:ADM}. PCADM results obtained for various wind–wave conditions are presented in Sec.~\ref{sec:PCADM}, followed by an analysis of phase effects associated with relative platform pitching motion in Sec.~\ref{sec:Phase}. Sec.~\ref{sec: wave and zo_lo} further isolates the contributions of wave phase and turbine pitch motion by comparing results from different representations of the waves via the bottom boundary condition used in LES. Finally, conclusions are given in Sec.~\ref{sec:Conc}. The Appendices include applications of PCADM to other turbines, and an error analysis focusing on statistical convergence of mean power trends obtained from simulations.

\section{\label{sec:wind-turbine} Power Curve Implementation of Actuator Disk Model}

The Actuator Disk Model (ADM) computes aerodynamic momentum and power extraction\cite{Burton2021WindEnergyHandbook} from the incoming flow due to a wind turbine represented by a rotor disk with area $A_{\rm d}$ placed in an inflow  with characteristic velocity $U_\infty$. For turbulent flows, such as flows in wind turbine arrays including wake interactions, the upstream velocity $U_\infty$ is challenging to evaluate in LES since it is not immediately clear at what upstream sampling distance to assume induction to be negligible. Therefore, many LES simulations \cite{Stevens_Meneveau_2017, Shapiro_Gayme_Meneveau_2018,Xiao_Yang_2019} follow the  local formulation of Calaf {\it et al.}\cite{Calaf_Meneveau_Meyers_2010}, which writes the thrust force in terms of the local thrust coefficient $C'_T$ and  disk averaged velocity $U_{\rm d}$ at the rotor disk location rather than far upstream. The corresponding thrust force is given by
\begin{equation}
     \mathbf{F}_\text{d}=- \frac{1}{2} \,\rho\,  C'_T \,  U_{\rm d}^2 \, A_\text{d}  \,\mathbf{\hat{n}}, \label{eq:forceADM}
\end{equation}
where $\rho$ is air density, $C'_T$ the local turbine thrust coefficient, $\hat{\bf n}$ is the unit vector normal to the turbine disk, oriented downstream. (i.e. if the flow is in the $x$-direction, for a non-moving fixed bottom turbine we have $\hat{\bf n} = (1,0,0)$), and $U_{\rm d}$ is the disk-averaged velocity normal to the disk. In principle, since the incoming air velocity can depend on position, $U_{\rm d}$ is defined as  
\begin{equation}
    U_{\rm d} =\iiint \mathcal{R}\mathbf{(x)} \, \mathbf{u}\mathbf{(x)} \cdot \mathbf{\hat{n}} \,d^3 \mathbf{x}, \label{eq: U_d^r}.
\end{equation}
where $\mathbf{u} ({\bf x})$ is the air velocity at the rotor disk. $\mathcal{R}\mathbf{(x)}$ is a normalized weighting factor (filter function) that is constant for grid points fully inside the disk, 0 for points fully outside, and for partially overlapping cells, it is proportional to the fraction of the cell area contained within the disk\cite{Meyers_Meneveau_2010_AIAA}.  In LES, the total force $\mathbf{F}_\text{d}$ is then distributed uniformly on the simulation grid points where the turbine disk is located \cite{Wu2011, Calaf_Meneveau_Meyers_2010}. In most applications of ADM, turbines are assumed to operate in Region 2 (between the cut-in and rated power velocities), where the induction factor, thrust coefficient, and power coefficient are held constant at their optimal or near-optimal values, independent of the instantaneous wind speed. Under ideal condition, the maximum power coefficient is given by the Betz limit, attained at $a=\frac{1}{3}$, for which $C_T=\frac{8}{9}$ and $C_P=\frac{16}{27}$. The corresponding local values are $C'_P = C'_T=2$. In wind-farm LES, turbines are typically modeled with constant coefficients corresponding to operating conditions below this theoretical optimum. We refer to this traditional formulation of the ADM as the constant-coefficient actuator disk model (CCADM), which has been widely used in previous studies\cite{Calaf_Meneveau_Meyers_2010,Stevens_Meneveau_2017,Narasimhan_Gayme_Meneveau_2022,Xiao_Yang_2019,Xiao_Zhu_Narasimhan_Gayme_Meneveau_2025,Kasper_Stieren_Stevens_2024}. For example, Calaf, et al\cite{Calaf_Meneveau_Meyers_2010} used $a=\frac{1}{4}$ and $C'_T=1.33$.

In this work we are interested in developing an actuator disk model that takes into account the full range of turbine operation. Turbine reference data usually includes curves specifying the thrust coefficient $C_T$ as function of the far-field inflow velocity $U_\infty$. Different operating regimes are specified depending on this $U_\infty$ (for example, separating Region 2 from 3 at some specified rated operation point speed $U_{\rm tr}$).  However, in LES where $U_{\rm d}$ is known but not $U_\infty$, their relationship (as well as the corresponding value of $C_T^\prime$) depends on the induction factor $a$ that also depends on possible tilting/yawing angles and the operating regime. Therefore, constructing an appropriate mapping between the manufacturer-provided $C_T(U_\infty)$ data and the LES variable $C_T^\prime(U_{\rm d})$ is a key step in the proposed PCADM formulation.

\begin{figure*}[h] 
\includegraphics[width=\columnwidth]{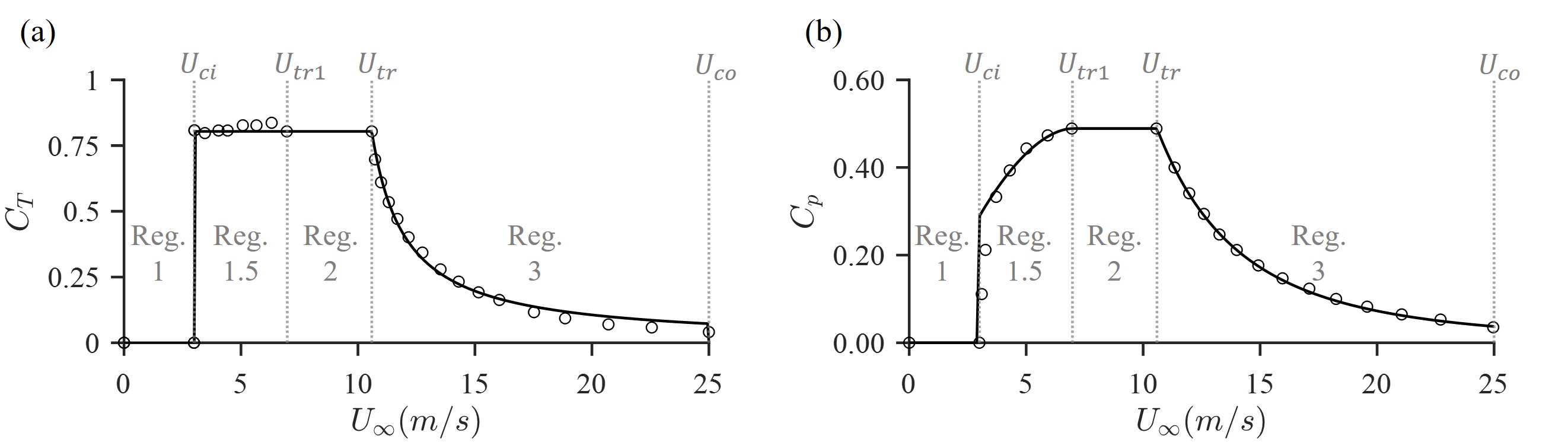}
\caption{Fitted thrust (a) and power (b) coefficients as functions of far upstream wind speed for  PCADM (solid lines), and the reference ROSCO data provided in Ref. \cite{IEA15MW_ROSCO} (circular markers). 
} 
\label{fig:ROSCO}
\end{figure*}

We begin by proposing a piecewise fitting function for the thrust-coefficient curve 
\begin{equation}
C_T(U_\infty) =
\begin{cases}
0, &  U_\infty \leq U_{\rm ci}\\
C_{\rm{T,op}}, & U_{\rm ci}<U_\infty \leq U_{\rm tr}\\[3pt]
\dfrac{C_{\rm T,op}}{1+\alpha(U_\infty/U_{\rm tr} - 1)}, & U_{\rm tr} < U_\infty \leq U_{\rm co}\\
0, &   U_{\rm co} < U_\infty.
\end{cases}
\label{eq:CT-U_inf}
\end{equation}
Here $C_{\rm T,op}$ denotes the operational thrust coefficient in Region 2. In this Region, where the disk-normal inflow velocity $U_\infty$ lies between cut-in speed ($U_{\rm ci}$) and the transition speed to Region 3 ($U_{\rm tr}$), the turbine operates at a constant thrust coefficient.  Region 3 begins when $U_\infty > U_{\rm tr}$ and continues until the inflow speed reaches the cutoff speed $U_{\rm co}$. For inflow conditions outside these operational bounds, the turbine is shut down and $C_T = 0$. The parameter $\alpha$ is a turbine-specific fitting coefficient. Eq.~(\ref{eq:CT-U_inf}) could easily be modified to account for turbines featuring Region 1.5 and Region 2.5 operation. For the IEA~15MW turbine, which is the focus of this study, Region 1.5 exhibits a nearly constant $C_T$ behavior similar to Region 2 and is therefore merged into Region 2 for simplicity. The resulting fitted curve for the PCADM  for the IEA~15MW turbine is presented in Fig.~\ref{fig:ROSCO}(a). For this turbine, the fitted parameters are 
\begin{equation} C_{\rm T,op} = 0.804, ~~~\alpha=7.413, ~~{\rm and}~~U_{\rm ci} = 3\mathrm{m/s},~~ U_{\rm tr} = 10.59\mathrm{m/s},~~ U_{\rm co} = 25\mathrm{m/s}.
\end{equation}
We remark that if we assume that the turbine is operating in Region 2, the corresponding local coefficient for a CCADM is $C_T'=1.545$. Fig.~\ref{fig:ROSCO} also includes the fitted power coefficient curve, to be discussed later, for which $U_{\rm tr1}=6.98 \mathrm{m/s}$. 

In order to obtain a useful $C_T^\prime-C_T$ relationship, we use the following formula for the induction factor $a$ as function of $C_T$ for a turbine is inclined (i.e., tilted) at an angle $\theta$ from Heck {\it et al.}
\cite{Heck_Johlas_Howland_2023}
\begin{equation}
a = \dfrac{1}{2}\left(1-\sqrt{1-C_T\cos{\theta}}\right).
\label{eq: a-U_inf}
\end{equation}
Since the actuator disk theory gives $U_\infty=U_{\rm d}/(1-a)$, the induction factor can be expressed as a function of the disk velocity $U_{\rm d}$. In Region 2, the thrust coefficient is $C_T=C_{\rm T,op}$. Therefore, Eq.~\ref{eq: a-U_inf} immediately yields a constant induction factor $a=\frac12\left(1-\sqrt{1-C_{\rm T,op}\cos{\theta}}\right)$, which is independent of $U_{\rm d}$. Consequently, when the far-field velocity reaches the Region 2--Region 3 transition point $U_\infty = U_{\rm tr}$, the corresponding transition disk velocity is
\begin{equation}
 U_{\rm d,tr} =  U_{\rm tr} \, \dfrac{1}{2}\left(1+\sqrt{1-C_{\rm T,op}\cos{\theta}}\right),
 \label{eq:udtr}
\end{equation}
which provides the equivalent transition criterion in terms of the LES-ready variable $U_{\rm d}$. It is based upon knowledge of the reference transition velocity $U_{\rm tr}$, the angle $\theta$, and the operating (optimal) thrust coefficient $C_{\rm T,op}$.

In Region 3, however, the thrust coefficient is no longer constant. Instead, the Region 3 branch of Eq.~\ref{eq:CT-U_inf} is substituted into Eq.~\ref{eq: a-U_inf}. Replacing the far-field velocity by the actuator-disk relation $U_\infty=U_{\rm d}/(1-a)$ produces an equation containing only the induction factor $a$. After simplification, this equation reduces to a quadratic equation in $a$. Selecting the physically admissible solution satisfying ($0<a<1/2$) yields 
\begin{equation}
a(U_{\rm d}, \theta) = \dfrac{1}{2}
\begin{cases}
0, &  U_{\rm d} \leq U_{\rm d,ci}\\
1-\sqrt{1-C_{\rm T,op}\cos{\theta}}, & U_{\rm d, ci}<U_{\rm d} \leq U_{\rm d,tr}\\[3pt]
1-\dfrac{\alpha}{\alpha-1}\dfrac{U_{\rm d}}{U_{\rm tr}}+\sqrt{\left( 1-\dfrac{\alpha}{\alpha-1}\dfrac{U_{\rm d}}{U_{\rm tr}}\right)^2+\dfrac{C_{\rm T,op}\cos{\theta}}{\alpha-1}}, & U_{\rm d,tr} < U_{\rm d} \leq U_{\rm d,co}\\
0, &   U_{\rm d,co} < U_{\rm d},
\end{cases}
\label{eq: a-U_d}
\end{equation}
which provides a continuous piecewise expression for the induction factor over all operating regions. Similar to $U_{\rm tr}-U_{\rm d,tr}$ mapping, the $U_{\rm ci}-U_{\rm d,ci}$  and $U_{\rm co}-U_{\rm d,co}$ relation at cut-in and cut-out velocity are given by $U_{\rm d,ci} =  \dfrac{U_{\rm ci}}{2} \, \left(1+\sqrt{1-C_{\rm T}\cos{\theta}}\right)$ and $U_{\rm d,co} =  \dfrac{U_{\rm co}}{2} \, \left(1+\sqrt{1-C_{\rm T}\cos{\theta}}\right)$, respectively. For the usual case ($\alpha>1$), Eq.~\ref{eq: a-U_d} always satisfies ($0<a<1/2$), indicating that the turbine remains in the windmill operating state and the classical actuator disk momentum theory remains valid throughout the conditions relevant to the present study.

\begin{figure}[h!] 
\includegraphics[width=0.8\columnwidth]{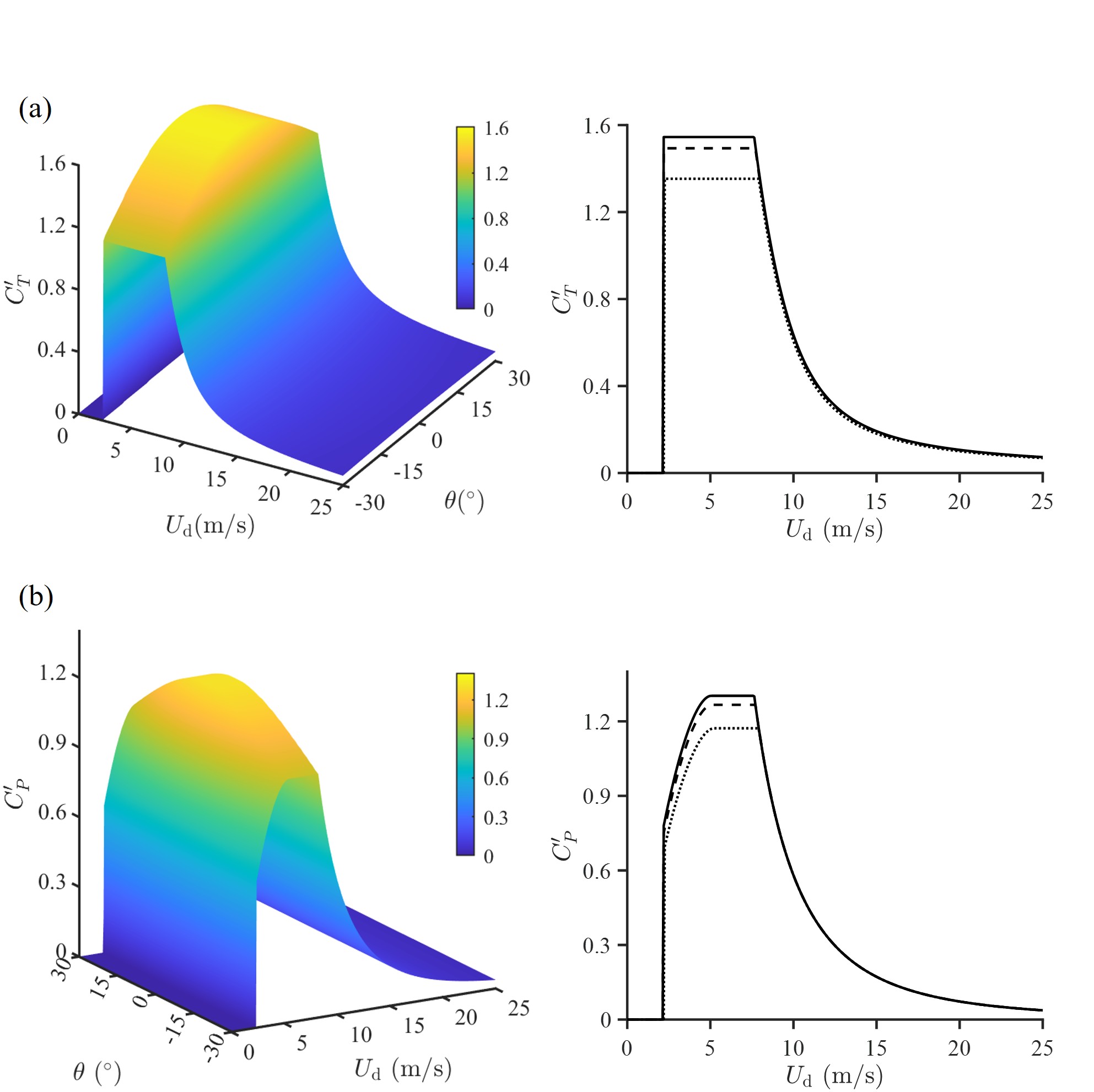}
\caption{Piecewise local coefficients for the IEA-15MW reference turbine using ROSCO\cite{IEA15MW_ROSCO}: (a) $C_T'$ and (b) $C_p'$. The left column displays the 3D response surface and the right column shows 2D curves at specific pitch angles ($\theta = 0^\circ$ solid line, $\theta = 10^\circ$ dashed line, and $\theta = 20^\circ$ dotted line). The fitting parameters are defined as $C_{\rm T,op}' = 1.545, ~~C_{\rm p,op}' = 1.281, ~~U_{\rm ci} = 3\mathrm{m/s},~~U_{\rm tr1} = 6.98\mathrm{m/s}, ~~ U_{\rm tr} = 10.59\mathrm{m/s},~~ U_{\rm co} = 25\mathrm{m/s}, ~~{\rm and}~~\alpha=7.413,~~ \gamma=0.55,~~\sigma=1.8$.
} 
\label{fig: CT_p}
\end{figure}

Finally, using the momentum-theory relation $C_T=4a(1-a)$, the local thrust coefficient required by the LES is obtained as $C_T'(U_{\rm d}, \theta) = C_T/(1-a)^2=4a/(1-a)$, where $a(U_{\rm d},\theta)$ is given by Eq.~\ref{eq: a-U_d}. The resulting $C_T'$ surface is shown in Fig.~\ref{fig: CT_p} (a). In Region 2, $C_T'$ remains independent of the disk velocity and decreases symmetrically on both sides of $\theta=0^\circ$. In Region 3, it decreases with increasing disk velocity while again displaying a symmetrical reduction on both sides away from $\theta=0^\circ$.   

Before discussing the treatment of tilting turbines and dependence on relative velocity, we present the corresponding fited function for the power-coefficient for the IEA 15 MW turbine. The power will be evaluated as usual\cite{Burton2021WindEnergyHandbook} according to
\begin{equation}
    P = \frac{1}{2}\rho \, C_P\, A_{\rm d} \, U_\infty^3 = \frac{1}{2}\rho \, C_P^\prime\, A_{\rm d} \, U_{\rm d}^3.
    \label{eq:powerADM}
\end{equation}
Based on the  $C_P$ curve provided for the IEA 15~MW turbine, the following piecewise representation as function of $U_\infty$ is obtained:
\begin{equation}
C_P(U_\infty)  =
\begin{cases}
0, &  U_\infty\leq U_{\rm ci}\\[3pt]
 C_{\rm P,op} - \gamma\,(1-U_\infty/U_{\rm tr1})^\sigma, & U_{\rm ci}<U_\infty \leq U_{\rm tr1}\\[3pt]
C_{\rm P,op}, & U_{\rm tr1}<U_\infty \leq U_{\rm tr}\\[3pt]
C_{\rm P,op}(\dfrac{U_{\rm tr}}{U_\infty})^3, & U_{\rm tr} < U_\infty \leq U_{\rm co}\\
0, &  U_{\rm co} < U_\infty.
\end{cases}
\label{eq: CP-U_inf}
\end{equation}
Here $C_{\rm P,op}$ represents the operational power coefficient in Region 2. The coefficient $C_P$ remains constant in Region 2 and decreases cubically with relative inflow velocity in Region 3 keeping the power constant at its nominal rated value $P_{\rm rt} = 15 \mathrm{MW}$ for this turbine. Unlike the $C_T$ curve, the $C_P$ curve of the IEA 15~MW turbine exhibits distinct behavior in Region 1.5, which is explicitly captured by the fitting parameters $\gamma$ and $\sigma$. The resulting fitted $C_P$ curve is presented in Fig.~\ref{fig:ROSCO}(b), with parameters $C_{\rm P,op} = 0.489$, $U_{\rm tr1} = 6.98\mathrm{m/s}$, $\gamma=0.55$, $\sigma=1.8$. Similarly, the relationship between $C_p^\prime$ and $C_p$ can be derived using the induction factor $a$, yielding $C_p'=C_p/(1-a)^3$. By replacing the far-field velocity with the disk velocity relation, $U_\infty=U_{\rm d}/(1-a)$, the resulting $C_p'$ surface is shown in Fig. 2 (b). In Region 1.5, $C_p'$ increases with increasing disk velocity $U_{\rm d}$. In Region 2, $C_p'$ remains nearly independent of $U_{\rm d}$. 
In Region 3, $C_p'$ decreases with increasing $U_{\rm d}$. In all cases, a symmetric decrease on both sides of $\theta=0^\circ$ is seen. 
The comparative study focuses on the IEA 15MW reference turbine, but the findings are expected to apply quite in general for floating turbine modeling, as demonstrated by the applications to the NREL 5 MW and IEA 22 MW turbines in Sec. \ref{sec:other turbines}.

\section{\label{sec:relativemotion} Relative motion for floating wind turbines} 
In the prior section we have developed a method to specify the local coefficients $C_T'$ and $C_p'$ based on the local velocity $U_{\rm d}$ for the case when $C_T$ and $C_p$ can  change for different operating regions. However, when the turbine is pitching due to floating conditions, the induced turbine motion and its pitch angle $\theta(t)$ will affect the relative inflow velocity to the actuator disk, affecting thrust and power extraction. Instead of $U_{\rm d}$, the velocity of the rotor that needs to be used in Eq.~\ref{eq:forceADM}  will  be the relative disk-averaged velocity. Throughout this study, we define a Cartesian coordinate system in which $x$ is the streamwise direction, $y$ is the spanwise direction, and $z$ is the vertical direction. The turbine pitches about an axis parallel to the $y$-direction, with its disk normal aligned with the $x$-direction when $\theta = 0^\circ$.  Placing the entire actuator disk momentum analysis in a frame moving with the turbine, and neglecting unsteady and non-inertial terms (see discussions in Refs. \cite{Dabiri2020Betz, Chattot2014ActuatorDisk}) the relevant air speed at the disk that generates aerodynamic forces is now given by  
\begin{equation}
    U_{\rm d}^{\rm rel}(\theta) =\iiint \mathcal{R}\mathbf{(x)}\Big(\mathbf{u}\mathbf{(x )}-{\bf v}_\text{d}\mathbf{(x)}\Big)\cdot \mathbf{\hat{n}}(\theta) \,d^3 \mathbf{x}, \label{eq:U_d^r}
\end{equation}
where $U_{\rm d}^{\rm rel}$ denotes the relative velocity averaged over the moving actuator disk at the disk location, and   ${\bf v}_\text{d}$ is the local disk velocity  
\begin{equation} 
{\bf v}_\text{d}(\mathbf{x}) = \frac{d  \mathbf{x}}{dt}  = r_\perp \,\frac{d   \theta}{dt} \, \hat{\mathbf{n}}(\theta).
\end{equation}
Here $\mathbf{x}$ is the position vector on the pitching actuator disk and $r_\perp$ is the corresponding radius of rotation, i.e.
$r_\perp=|{\bf r} - ({\bf r}\cdot{\bf e}_y)\, {\bf e}_y|$ with
${\bf r}={\bf x}-{\bf x}_{\rm plat}$ being the position relative to ${\bf x}_{\rm plat}$, the platform center of rotation, and ${\bf e}_y$ being the unit vector in the direction of rotation 
(assumed to be in the $y$-direction for non-heaving turbines).  Also, $\hat{\bf n}(\theta)$ is the disk-normal unit vector which depends on pitch angle $\theta$ according to
\begin{equation}
\hat{\mathbf{n}}(\theta) = \mathbf{R}_y(\theta) \cdot \hat{\mathbf{n}}_0,~~~~~ {\rm with}~~~~~~~
\mathbf{R}_y(\theta) =
\begin{bmatrix}
\cos\theta & 0 & \sin\theta \\
0 & 1 & 0 \\
-\sin\theta & 0 & \cos\theta
\end{bmatrix},
\end{equation}
and $\hat{\bf n}_0=(1,0,0)$ for flow and non-tilting (resting) turbine's unit normal in the $x$-direction. The filtering function $\mathcal{R}\mathbf{(x)}$ is assumed to be moving with the disk as well. The time-varying angle $\theta=\theta(t)$ will be prescribed based on floating platform motion responding to a prescribed ocean wave field.

In the moving turbine-attached frame, the entire analysis performed in the prior section is still applicable, if we neglect additional terms in the momentum equation coming from unsteadiness and non-inertial conditions. We also must assume that the thrust and power curves provided for the turbine that have been determined under stationary conditions in an inertial frame are still applicable even in the nominally accelerating frame of a floating turbine. These  assumptions are necessary in practice, given the absence of reliable turbine data under time-varying conditions and in non-inertial frames of reference. Therefore, we assume that a Galilean transformation of the velocity field does not alter the relationship between $U_{\rm d}$ and $U_\infty$, and that the induction factor based relations derived in the previous section remain valid.

\subsection{Temporal filtering}
The above derivations assumed that the turbine responds instantaneously to incoming velocity, leading to the 
``steady-state'' performance commonly assumed for reference wind turbines \cite{Jonkman_et_al_2009, Bak_et_al_2013, gaertner2020iea15mw} and illustrated in Fig.~\ref{fig:ROSCO}. In practice, however, the OFWF is a dynamic system, and the rotor cannot adjust rapidly enough to follow high-frequency relative incoming velocity fluctuations because of inertia and limitations in dynamic response. Various control strategies are often applied in practice to smooth the turbine response, mitigating excessive structural loads and fatigue while preventing possible control instabilities. In order to mimic  these additional smoothing effects in LES we apply a one-side exponential temporal filter to the relative disk-averaged velocity, $U_{\rm d}$ according to
\begin{equation}
U_{d,\tau}^{\rm rel}(t) = \frac{1}{\tau} \int_{-\infty}^{t} U_{\rm d}^{\rm rel}(t') 
\, e^{-(t - t')/\tau} \, dt'.
\label{eq:Filt_U_d}
\end{equation}

In simulations, this filtering is implemented using the updates  $U_{d,\tau}(t) = U_{d,\tau}(t-\delta t)(1-\epsilon) + \epsilon \, U_{\rm d}(t-\delta t)$ where $\epsilon = \delta t/\tau$, $\delta t$ is the simulation time-step., 
The filter time scale $\tau$ is meant to reflect the turbine’s physical response limits imposed by its control system. In standard reference wind turbine control, generator torque control governs operation in Region 1.5 and Region 2, where the torque rate limit sets the dynamic response. Above the rated wind speed ($U_{\rm tr}$), collective blade pitch control dominates (Region 3), with the blade pitch rate limit becoming the primary constraint. For the IEA~15MW turbine, the maximum blade pitch rate is $0.035\mathrm{rad/s}$\cite{IEA15MW_ROSCO} corresponding to a characteristic time scale of $\tau \approx 4.57 \mathrm{s}$. In this study the filter time scale is set to be $\tau = 4.5\mathrm{s}$ across all regions. 

From here onward, the turbine thrust and power 
are determined using Eqs. \ref{eq:forceADM} and \ref{eq:powerADM}. The relative input velocity $U_{\rm d}^{\rm rel}$ is obtained from Eq.~\ref{eq:U_d^r} using the time filtered local disk velocity $U_{d,\tau}^{\rm rel}$ (Eq. \ref{eq:Filt_U_d}). 

\subsection{\label{sec:Motion}Platform motion and turbine tilting angle $\theta(t)$}

Simple monochromatic deep-water waves propagating in the streamwise ($x$) direction are considered, characterized by wave amplitude $A$, angular wave frequency $\omega$, and wavenumber $k$. The wave surface elevation given by $\eta(x,t) = A\cos(kx-\omega t)$. The wavelength and wave period are $\lambda=\frac{2\pi}{k}$ and $T=\frac{2\pi}{\omega}$, respectively. Under the deep water assumption, the dispersion relation reduces to $\omega^2=gk\tanh (2Ak) \approx gk$, such that the wave frequency is uniquely determined by the prescribed wavenumber or wavelength. Consequently, the wave phase speed $c=\omega/k$ depends on the wavelength and is independent of the water depth.

\begin{figure} [h!]
\centering
\includegraphics[width=0.8\columnwidth]{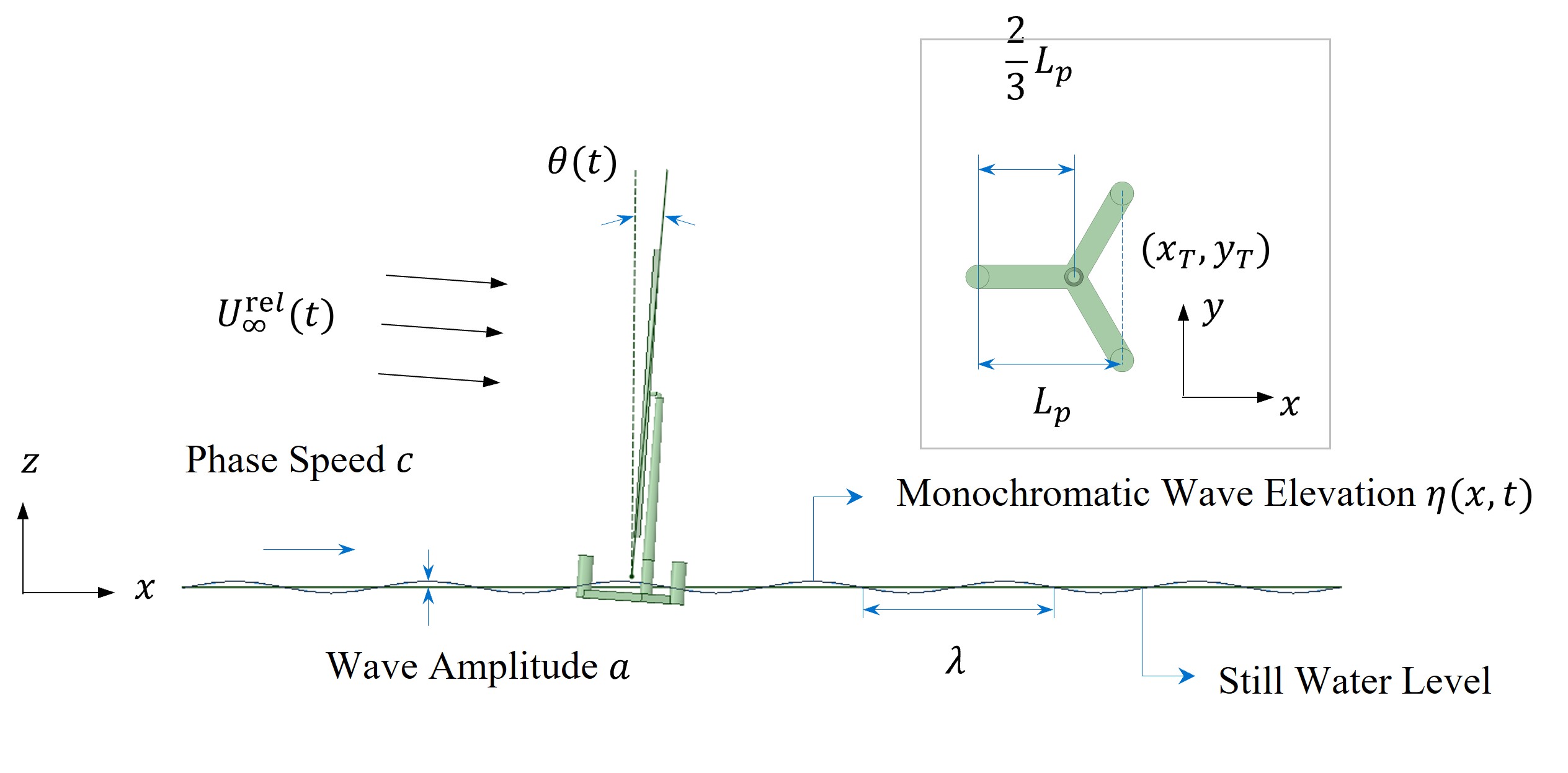}
\caption{Schematic of the pitching motion of an individual turbine} 
\label{fig:Schematic}
\end{figure}

The University of Maine (UMaine) reference floating offshore wind turbine semi-submersible VolturnUS-S platform\cite{Allen_et_al_UMaine_VolturnUS_15MW} is designed to support the IEA-15MW reference wind turbine, and is therefore used in this study. The hull consists of three buoyant columns, each with a diameter of 12.5 m, arranged radially such that their centers are located 51.75 m from the tower’s vertical axis. The characteristic platform length is defined as the distance from one buoyant column of the triangular arrangement to the opposite edge connecting the other two columns such that $L_P=90\text{m}$, see Fig.~\ref{fig:Schematic}. 

We assume the pitching turbine motion  responds to the simple monochromatic deep-water wave, and the floating columns follow the water surface elevation at all times, as illustrated in Fig. \ref{fig:Schematic}. Therefore, for a turbine located at $x_T$, the tangent of the platform pitching angle $\theta$ can be related to the ratio between the leading and the trailing edge of the platform and the platform length, $L_P$ according to
\begin{eqnarray}
\tan\theta(t)= \frac{1}{L_p} \left(\eta(x_T-\frac{2L_p}{3},t)-\eta(x_T+\frac{L_p}{3}, t)\right) 
 =\frac{2A}{L_P}\sin(\frac{L_P}{\lambda})\sin(kx_T-\omega t-\frac{k}{6}L_P),
\label{eq: Pitching}
\end{eqnarray}
where it was assumed that the turbine location $x_T$ is at the geometric midpoint of the platform, i.e. at distances $2L_p/3$ and $2L_p/3\cos(60^\circ)$ in the $x$-direction from each column (see Fig. \ref{fig:Schematic}).
The pitching motion therefore depends on the turbine location, $x_T$, wave characteristics, i.e., $a, \lambda, k, \omega$, as well as the platform length $L_P$. We assume that the pitching motion frequency is the same as the passing wave frequency. 

The relative pitching phase between turbines in a wind farm depends on the ratio between the wave wavelength and the turbine spacing. When downstream turbine separation is an integer multiple of the wavelength, their pitching motions are in-phase (IP), such that all turbines are simultaneously at the same pitching angle. When the spacing corresponds to an odd multiple of half the wavelength, turbines directly downstream pitch in opposite-phases (OP), with neighboring turbines exhibiting pitching angles of opposite sign. If the relation  between turbine spacing and wavelength does not follow a simple relationship (e.g. is not a simple rational number) the  pitching motions are termed to be of incommensurate-phase (IC).

 \section{\label{sec:LES}LES Methodology and simulation setup}

\subsection{Governing equations and LES code}
We consider a fully developed, neutral stability, turbulent boundary layer  flow forced in the streamwise ($x$) direction by a pressure gradient. We perform LES with the JHU code, LESGO which has been extensively validated and can accurately simulate the ABL with wind farms \cite{Calaf_Meneveau_Meyers_2010, Xiao_Yang_2019,Xiao_Zhu_Narasimhan_Gayme_Meneveau_2025}. LESGO solves the LES filtered incompressible Navier-Stokes (NS) equations using a pseudospectral discretization in the horizontal directions and second-order centered finite differencing in the wall-normal direction. 
\begin{eqnarray}
    \frac{\partial \tilde{u}_i}{\partial x_i} && = 0   \\
    \frac{\partial \tilde{u}_i}{\partial t} +\tilde{u}_j\frac{\partial\tilde{u}_i}{\partial x_j} &&=-\frac{1}{\rho}\frac{\partial\tilde{p}^*}{\partial x_i}- \frac{\partial \tau_{ij}^{d}}{\partial x_j} 
    - \frac{\delta_{i1}}{\rho}\frac{\partial p_{\infty,c}}{\partial x_1} +f_i \label{eq: NS equation}
\end{eqnarray}
where $\tilde{u}_i$ is the filtered velocity field, $\tilde{p}^*=\frac{\tilde{p}}{\rho}+\frac{\tau_{kk}}{3}$ is the filtered modified pressure, and $\tau_{ij}^{d}$ is the deviatoric part of the subgrid-scale (SGS) stress tensor $\tau_{ij}$. It is modeled using a scale-dependent Lagrangian dynamic model \cite{BouZeid_Meneveau_Parlange_2005} and the trace of the SGS stress tensor ($\tau_{kk}$) is combined into the modified pressure as is usually done for LES of incompressible flow. $f_i$ is a modeled wind turbine thrust force (per unit mass) and acts as a momentum sink in Eq.~(\ref{eq: NS equation}). It is nonzero only in regions where wind-turbine rotor disks are located (See Section \ref{sec:wind-turbine} for more details on the wind turbine thrust force calculation). 
Temporal integration is performed using the explicit second-order Adams-Bashforth method. The term $\partial p_{\infty,c}/\partial x_1$ is used to maintain the velocity at top of the domain to a specified value using a PI controler, described in the next section.

\subsection{\label{sec:pressure}Pressure gradient controller}

 Wind in the ABL are driven by the geostrophic wind, and the outer layer is influenced by Coriolis accelerations\cite{Tennekes_Lumley_1972}.
In studies of very large wind farms, fully developed conditions can be assumed for which the  large-scale forcing is commonly parameterized through an imposed constant streamwise pressure gradient, while neglecting Coriolis effects\cite{Calaf_Meneveau_Meyers_2010}. In this formulation, the flow is driven jointly by the prescribed pressure gradient and by shear stresses (momentum exchange) at the surface and within the turbine region, resulting in a coupled adjustment of the  wind velocity above and within the wind farm. In order to provide a common reference condition above the wind farm, in the present study we employ a control algorithm that maintains a constant velocity at the top of the domain as a prescribed input parameter. As a result, the required pressure gradient, $\partial  p_{\infty,c} / \partial x_1$, adjusts dynamically in time to satisfy the imposed desired reference velocity following 
\begin{equation}
    -\frac{1}{\rho}\frac{\partial\langle p_\infty \rangle_c}{\partial x_1}(t+\Delta t)= -\frac{1}{\rho}\frac{\partial\langle p_\infty \rangle_c}{\partial x_1}(t)+K_pe(t)+K_i\int_0^te(t')dt'.\label{eq: PI-Controller}
\end{equation}
The last two terms represent the control signal (the difference between the pressure gradient at the current time step and the previous time step). This PI controller has a dimensionless  proportional gain $K_p$ (set to $K_p=0.5$), and an integral gain $K_i$  (set to $K_i = 10^{-4}\,\mathrm{s}^{-1} $). The error signal is defined as
\begin{equation}
    e(t)=\frac{1}{L_z}\,|U_{\rm ref} -\langle u_{\rm LES}(L_z)\rangle| \,\,(U_{\rm ref} -\langle u_{\rm LES}(L_z))\rangle), \label{eq: PI-error}
\end{equation}
where $\langle u_{\rm LES}(L_z) \rangle$ is  the planar averaged LES velocity at the top (at $z=L_z$) of the domain and the desired imposed velocity $U_{\rm ref}$. We chose this form so that $e(t)$ has the same physical units as the streamwise pressure gradient  (quadratically nonlinear).

\subsection{\label{sec: Wave} Wave-field representation in LES}

To predict the momentum flux at the moving wave surface, a phase-dependent moving surface drag (MOSD) wall model \cite{Ayala_et_al_2024_MOSD} is employed.  For selected cases as comparison, an equilibrium wall model (EQWM) \cite{Moeng_1984} based on Monin–Obukhov similarity theory (MOST) for rough surfaces is also applied. This comparison enables us to isolate  wave-phase-dependent effects from those induced by turbine motion, thereby quantifying the relative importance of the wave phase and the turbine pitching phase.

The MOSD model consists of two components: (i) a windward potential flow stress (WPM), based on ideal potential flow theory applied to windward-facing portions of the surface to represent form drag induced by flow blockage from horizontally resolved wave features, and (ii) an EQWM that follows MOST to account for the effects of unresolved surface roughness features. Stress predicted by the WPM is wave-phase dependent, as it evaluates the potential-flow pressure distribution on the windward-facing side of the waves, which varies and moves in space with the instantaneous wave motion. A modified EQWM is used to calculate the small scale interactions with the unresolved subgrid scale roughness features at the air-sea interface. A roughness length $z_{0,rp}$ is specified to represent drag from ripples and other unresolved small-scale features. Detailed information on MOSD model can be found in Ayala {\it et al.} \cite{Ayala_et_al_2024_MOSD}.    

In the classical equilibrium wall model (EQWM), the friction velocity $u^*$ from a wind profile near the surface can be determined using the surface roughness $z_0$ and the von Karman constant $\kappa$\cite{Tennekes_Lumley_1972}. In this model the wall stress $u_*^2$ is determined from the velocity at some grid-point above the surface based on assuming that the velocity follows the logarithmic law. However, a roughness length $z_0$ must be prescribed a-priori. Various methods exist to specify this parameter, see for instance the Charnock\cite{Charnock1955} and SWARL\cite{Ayala2026waveRoughness} relation for air-sea interaction, similarity-theory-based formulations\cite{MoninObukhov1954}, and equivalent sand-grain roughness models used in CFD wall functions\cite{Ansys2025}. Here we obtain $z_0$ through a fit of the logarithmic law from MOSD simulations with the same configuration, as detailed in Sec. \ref{sec: wave and zo_lo}

\subsection{\label{sec:Setup}Large Eddy Simulation Setup}

\begin{figure}[h!]
    \centering
    \includegraphics[width=0.9\columnwidth]{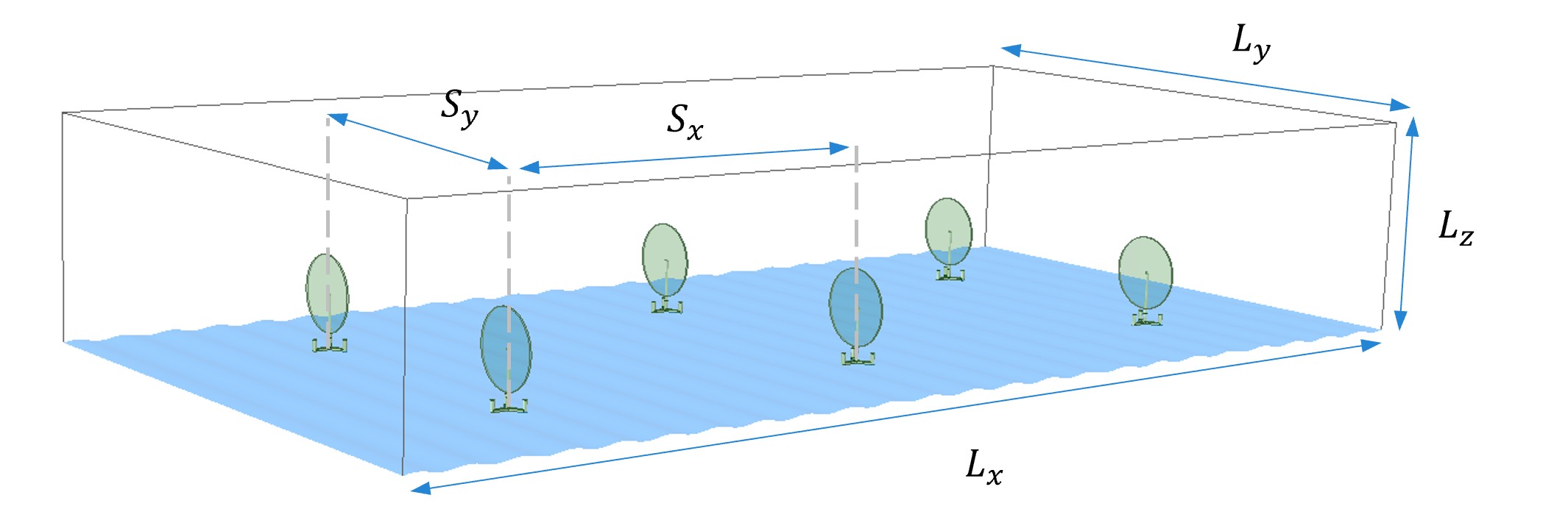}
    \caption{Schematic illustration of the  array of floating turbines in the computational domain of size $L_x\times L_y\times L_z$, subjected to periodic conditions in the $x$ and $y$ (horizontal) directions.}
    \label{fig: Computational-Domain}
\end{figure}

Simulations of flow in a fully developed OFWF are conducted within a computational
domain of dimensions $L_x \times L_y \times L_z = 3{,}840\mathrm{m} \times 1{,}920\mathrm{m} \times 700\mathrm{m}$. A schematic representation of the computational domain is shown in Fig.~\ref{fig: Computational-Domain}. Periodic boundary conditions are imposed in the streamwise ($x$) and spanwise ($y$) directions, while a stress-free condition is applied at the top boundary. At the bottom, both the EQWM and MOSD described in Sec.~\ref{sec: Wave} are employed. The  PI controller described in Sec.~\ref{sec:pressure} is applied to the pressure gradient forcing to maintain a desired reference velocity $U_{\rm ref}$ at the top of the domain. The wind farm consists of six IEA 15MW turbines arranged in two columns and three rows (Fig.~\ref{fig: Computational-Domain}). Each turbine has a rotor diameter of 240 m and a hub height of 150 m. By adjusting the ratio of wavelength $\lambda$ to turbine spacing $S_x$, different relative turbine pitching phases are obtained, namely IP, OP and IC  as described in  Sec.~\ref{sec:Motion}.  We also consider cases with stationary (fixed bottom) turbines (i.e., $\theta(t)\equiv 0^\circ$ subjected to the same wave conditions for comparison, we refer to these as FB cases. A summary of all simulated cases is provided in Table~\ref{tab:setup}. In order to ensure that each wave is resolved with at least eight grid points (for MOSD implementation), most of the simulations use a 
grid of $N_x \times N_y \times N_z = 192 \times 96 \times 128$, yielding $\Delta x \times \Delta y \times \Delta z = 20\mathrm{m} \times 20\mathrm{m} \times 5.4\mathrm{m}$. For smaller wavelength cases (FB-SW-12 and OP-SW-12), we use finer grids with $N_x \times N_y \times N_z = 384 \times 192 \times 256$, corresponding to grid spacings of $\Delta x \times \Delta y \times \Delta z = 10\mathrm{m} \times 10\mathrm{m} \times 2.7\mathrm{m}$.

\begin{table*}
\caption{\label{tab:setup}Summary of simulation parameters for the different LES cases.}
\begin{ruledtabular}
\begin{threeparttable}
\begin{tabular}{c c c c c c c c}
Case
& $U_{\text{top}}$ (m/s)
& $\lambda/S_x$
& $\theta_{\max}\;(^\circ)$
& $A$ (m)
& $ak$ 
& $T$ (s)
& $c$ (m/s)
\\
\hline
FB-15
& 15
& $1/8.5$
& 0
& 4.8
& 0.2
& 9.82
& 15.33
\\

OP-15
& 15
& $1/8.5$
& 5.8
& 4.8
& 0.2
& 9.82
& 15.33
\\

FB-C-15 \tnote{*}
& 15
& $1/8.5$
& 0
& 4.8
& 0.2
& 9.82
& 15.33
\\

OP-C-15 \tnote{*}
& 15
& $1/8.5$
& 5.8
& 4.8
& 0.2
& 9.82
& 15.33
\\

IP-15
& 15
& $1/8$
& 6.3
& 5.1
& 0.2
& 10.12
& 15.81
\\

IC-15
& 15
& $1/6\sqrt{2}$
& 5.8
& 4.8
& 0.2
& 9.83
& 15.35
\\

FB-EQ-15 \tnote{+}
& 15
& $1/8.5$
& 0
& 4.8
& 0.2
& 9.82
& --
\\

OP-EQ-15 \tnote{+}
& 15
& $1/8.5$
& 5.8
& 4.8
& 0.2
& 9.82
& --
\\

FB-10
& 10
& $1/8.5$
& 0
& 3.6 
& 0.15
& 9.82
& 15.33
\\

OP-10
& 10
& $1/8.5$
& 4.4
& 3.6
& 0.15
& 9.82
& 15.33
\\

IP-10 
& 10
& $1/8$
& 4.8
& 3.8
& 0.15
& 10.12
& 15.81
\\

IC-10
& 10
& $1/6\sqrt{2}$
& 4.4
& 3.6
& 0.15
& 9.83
& 15.35
\\

FB-12 
& 12
& $1/8.5$
& 0
& 4.8
& 0.2
& 9.82
& 15.33
\\

OP-12
& 12
& $1/8.5$
& 5.8
& 4.8
& 0.2
& 9.82
& 15.33
\\

FB-EQ-12 \tnote{+}
& 12
& $1/8.5$
& 0
& 4.8
& 0.2
& 9.82
& --
\\

OP-EQ-12 \tnote{+}
& 12
& $1/8.5$
& 5.8
& 4.8
& 0.2
& 9.82
& --
\\

FB-18 
& 18
& $1/8.5$
& 0
& 4.8
& 0.2
& 9.82
& 15.33
\\

OP-18
& 18
& $1/8.5$
& 5.8
& 4.8
& 0.2
& 9.82
& 15.33
\\

FB-SW-12 
& 12
& $1/15.5$
& 0
& 2.6
& 0.2
& 7.27
& 11.35
\\

OP-SW-12
& 12
& $1/15.5$
& 0.9
& 2.6
& 0.2
& 7.27
& 11.35
\\
\end{tabular}
\begin{tablenotes}
\item[*] Cases FB-C-15 and OP-C-15 are simulated using constant coefficient ADM. 
\item[+] Cases FB-EQ-15, OP-EQ-15, FB-EQ-12 and OP-EQ-12 are simulated using the EQWM, where the roughness height is obtained from the corresponding MOSD simulations.
\end{tablenotes}

\end{threeparttable}
\end{ruledtabular}
\end{table*}

\newpage

\section{Results}
A convergence analysis provided in Appendix~\ref{sec:convergence} shows that the variability among independent LES trials is smaller than the differences between the mean values of the cases considered. This analysis therefore supports the statistical robustness and reliability of the reported trends. 

\subsection{\label{sec:ADM} Documenting Effects of Power Curve ADM Implementation}

We first illustrate the impact of incorporating the turbine power curve into the ADM on the predicted power output and disk-averaged velocity through comparisons with the traditional CCADM. 

The analysis is performed for the $U_{\rm ref} = 15$ m/s cases, comparing FB-15 with FB-C-15 and OP-15 with OP-C-15. We examine the relative disk-averaged velocity $U_{\rm d}^{\rm rel}(t)$ and power generated at one of the turbines (due to periodic boundary conditions they are all statistically equivalent). Power is computed using $P = \frac{1}{2}\rho \, C_P^\prime\, A_{\rm d} \, U_{\rm d}^3$. The CCADM framework assumes a constant local power coefficient of $C_P'=1.281$, which is representative of Region 2 operation and evaluated using the $C_p'-C_p$ mapping at $\theta = 0$ and $C_P = C_{p, \rm op}=0.489$ (Fig. \ref{fig: CT_p}(b)). In contrast, the PCADM framework utilizes the disk-averaged velocity and pitch dependent mapping $C_p'(U_{\rm d},\theta)$ obtained from the local power coefficient curve illustrated in Fig. \ref{fig: CT_p}(b).

\begin{figure}[h!]
    \centering
    \includegraphics[width=0.9\columnwidth]{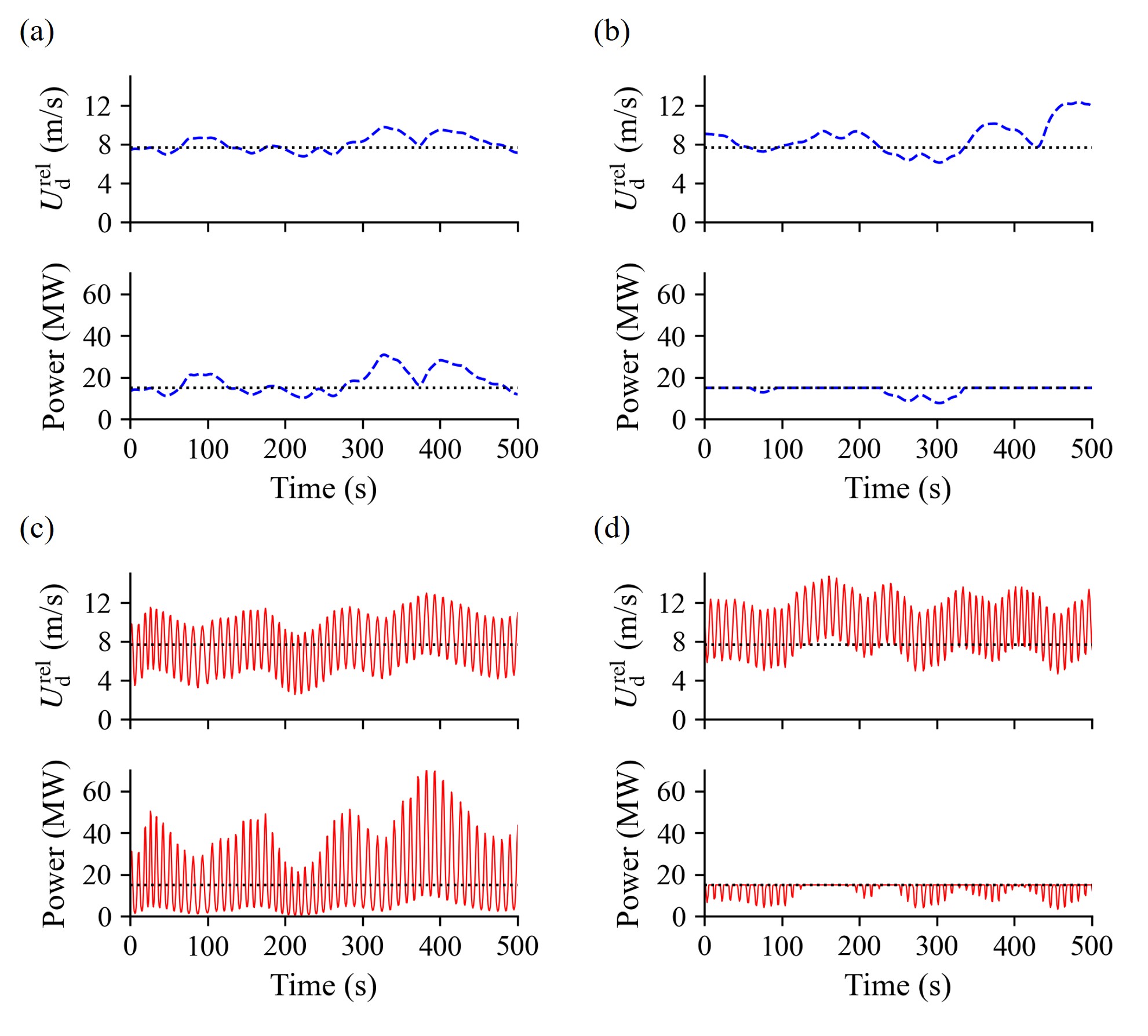}
    \caption{Sample 500 second time series of relative disk-averaged velocity and representative turbine power for fixed bottom cases, where results using CCADM (Case FB-C-15) are shown in (a), and using PCADM (Case FB-15) are shown in (b). Results for floating turbines (opposite phase) are shown in (c) for CCADM (Case OP-C-15) and in (d) for PCADM (Case OP-15). Black dotted lines represent the transition velocity $U_{\rm d, tr}$ between Region 2 and Region 3}
    \label{fig: ADM Series}
\end{figure}

Fig.~\ref{fig: ADM Series} shows representative time signals of  of relative disk-averaged velocity, $U_{\rm d}^{\rm rel}(t)$ and power output $P(t)$, for a sample 500 seconds time interval from each of the four simulations (FB-15, OP-15, FB-C-15, OP-C-15). Compared to the FB cases (panels (a) and (b) in Fig. \ref{fig: ADM Series}), which show large-scale turbulence-induced fluctuations occurring over time-scales ranging from tens to hundreds of seconds, pronounced periodic amplitude oscillations in velocity and power are observed for the floating turbine cases (panels (c) and (d)). These oscillations are caused by the wave-induced periodicity of the floating turbine pitching motion, causing a periodic modulation of the relative inflow velocity. At this reference velocity, the turbine operation region transitions frequently between Regions 2 and 3. The power output for the CCADM modeled turbine in Fig.~\ref{fig: ADM Series} (a) and (c) show unrealistically large power generation (i.e. far exceeding the rated condition of 15 MW),  when the turbine should be operating in Region 3 (i.e., when $U_{\rm d}>U_{\rm d,tr}$). Therefore, LES with the CCADM turbine model produces unrealistically high apparent power levels, indicating that failing to account for the region of operation in comparisons between floating and fixed-bottom turbines can be misleading. In contrast, the PCADM implementation in LES constrains the power output to its rated value when operating in Region 3 (see Fig.~\ref{fig: ADM Series} (b) and (d)). 

\begin{figure}[h!]
    \centering
    \includegraphics[width=0.6\columnwidth]{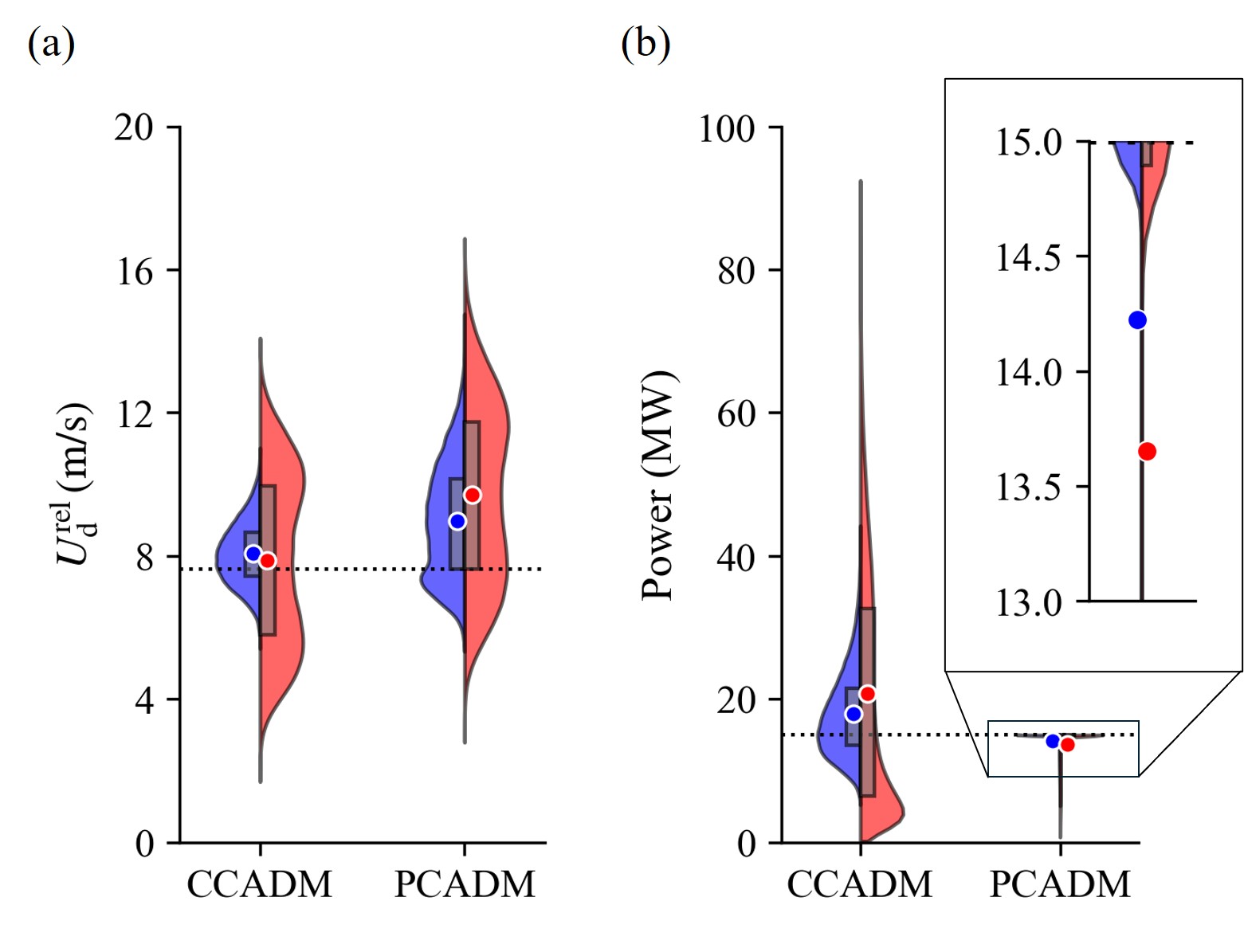}
    \caption{Probability density functions (violin plots) for (a) relative disk-averaged normal velocity defined in Eq.~\ref{eq:Filt_U_d} and (b) turbine power, comparing  classic constant $C_T'$ and $C_p'$ with power-curve aware (PCADM) methods. Blue (left sides) show results for fixed bottom (FB) while red (right sides) show results for floating (opposite phase) cases. The power results for PCADM on the right include an enlarged insert for clarity. The four cases correspond to FB-C-15, OP-C-15, FB-15, and OP-15 (see table \ref{tab:setup}).}
    \label{fig: ADM Violin}
\end{figure}

The probability density distributions (PDFs) of relative disk-averaged velocity and generated power for each of these cases are shown in the violin plots in Fig.~\ref{fig: ADM Violin}. These  also report the interquartile ranges, together with the mean values. For CCADM, the FB case exhibits an approximately Gaussian distribution of $U_{\rm d}^{\rm rel}$, whereas the OP case develops a bimodal distribution as a result of the periodic platform motion. When PCADM is employed, the FB case exhibits a narrow bump in the PDF immediately below the transition velocity $U_{\rm d,tr}$, indicating that the system tends to linger near the onset of Region~3 operation due to the change in the aerodynamic feedback of the rotor disk, for instance the thrust coefficients $C_T$ and $C_T'$, as well as the induction factor $a$. In contrast, the OP-PCADM case does not exhibit a pronounced feature near $U_{\rm d,tr}$, as the additional velocity fluctuations induced by the platform motion smooth out this localized accumulation of samples.

Consistent with trends already visible in Fig. \ref{fig: ADM Series}, the conclusion whether floating or fixed bottom turbines generate more mean power depends on whether one uses CCADM or PCADM to represent the turbine operation in LES. When using CCADM,  the predicted mean power generated by floating turbines is larger than for FB, while when using PCADM the reverse is true. The floating case yields lower mean power in this case due to more Region 2 operation than the FB case, which remains above rated conditions for most of this simulation time. 

\begin{figure}[h!]
    \centering
    \includegraphics[width=0.8\columnwidth]{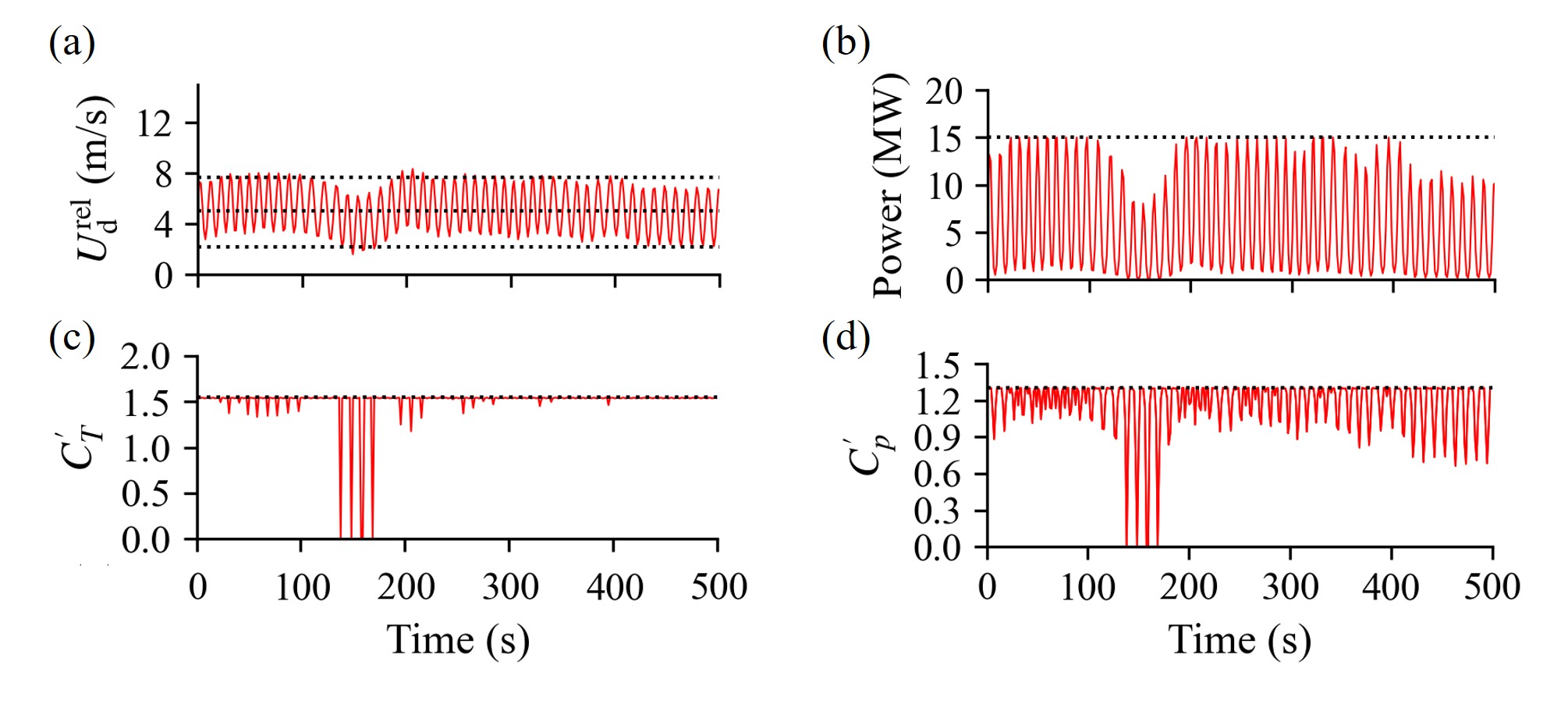}
    \caption{
    Sample 500 second time series of disk-averaged relative velocity (a) and turbine power (b) of a representative turbine in the array, for the case OP-10 at $U_{\rm ref} = 10$ m/s, using PCADM. Panels (c) and (d) show the respective thrust and power coefficient time-series.}
    \label{fig: PCADM serise}
\end{figure}

Next, we document the model behavior at a lower reference velocity $U_{\rm ref} = 10$ m/s, in this case the turbines are expected to operate mostly in Region 2. 
Fig.~\ref{fig: PCADM serise} shows signals of the relative disk-averaged velocity $U_{\rm d}^{\rm rel}$ and the power output for a representative turbine (Case OP-10). Also shown are signals of the instantaneous $C_T'$ and $C_p'$ coefficients for this case. As expected, under this wind-wave condition, the floating wind turbine stays mostly in Region 2, with occasional excursions into Regions 1, 1.5, and 3. 
PCADM captures the corresponding changes in $C_T'$ and $C_p'$ in a physically consistent manner. Specifically, $C_T'$ remains constant in Regions 1.5 and 2, with a slight reduction observed in Region 3. Similarly, $C_p'$ remains constant in Region 2 but decreases in Region 1.5. Both $C_T$ and $C_p$ drop to zero when the turbine operates in Region 1 (from approximately 140-180s in Fig.~\ref{fig: PCADM serise}). 

These results demonstrate the importance of accounting for region-dependent behavior. Therefore only PCADM is employed in the following analysis to evaluate power output under different wind–wave conditions and relative phase differences.

\subsection{\label{sec:PCADM} Pitching effect on turbine velocity and power for different wind and wave conditions}

In this section, we consider results at four reference wind conditions 
$U_{\rm ref} =10\mathrm{m/s}$,  $12\mathrm{m/s}$,  $15\mathrm{m/s}$ and  $18\mathrm{m/s}$ for a wavelength of  $\lambda = S_x/8.5$. We also examine the effects of wavelength for the intermediate case at $U_{\rm ref} =12\mathrm{m/s}$, where we include results at a shorter wavelengths $\lambda = S_x/15.5$ (denoted as cases FB-SW-12 and OP-SW-12). Simulations are performed for both FB and OP turbines to examine the influence of platform pitching. The power output for each case is analyzed to assess whether floating turbine motion enhances or reduces power generation and how the trends may depend on reference wind velocity and ocean wavelength. 

\begin{figure}[h!]
    \centering
    \includegraphics[width=0.67\columnwidth]{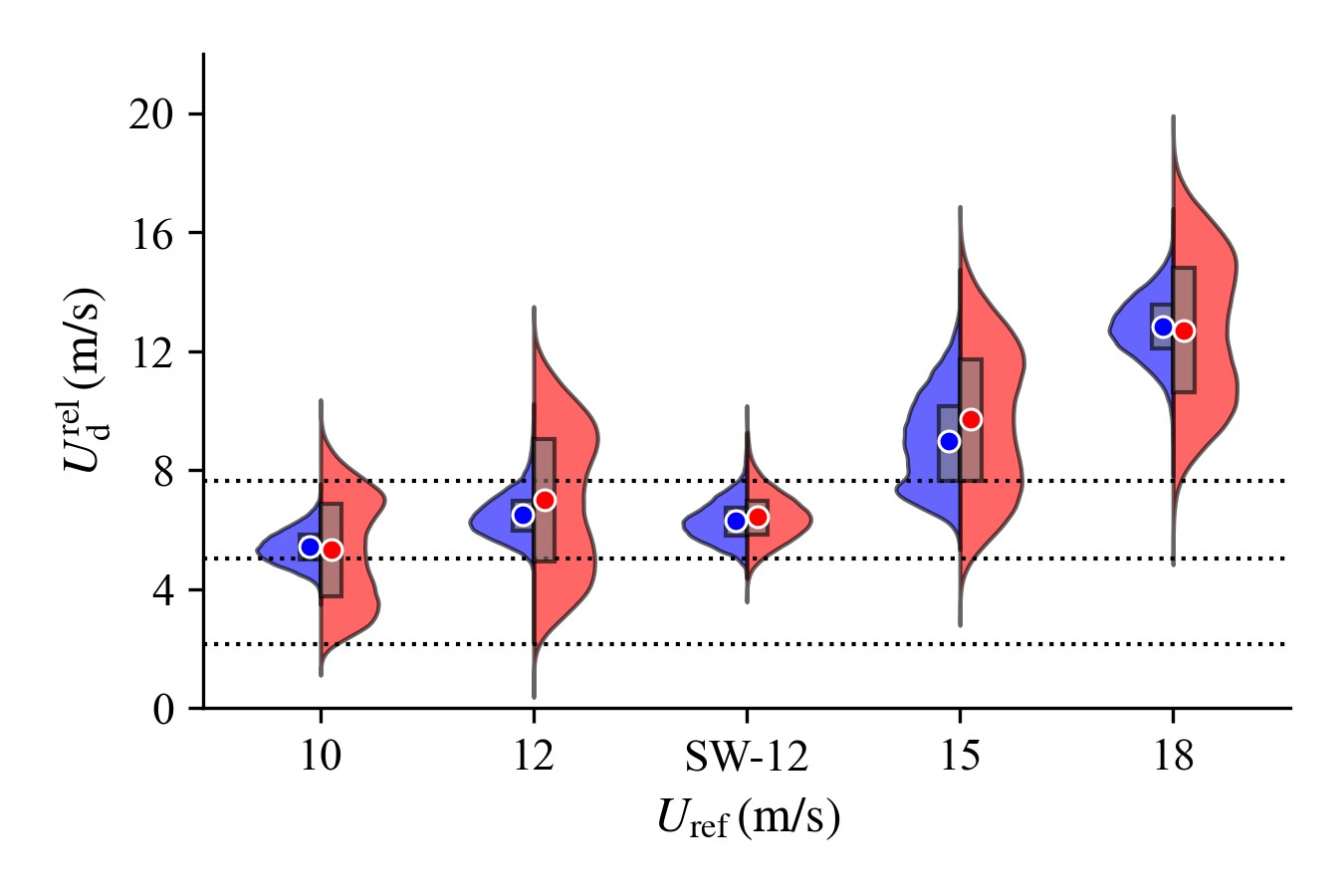}
    \caption{Violin plots showing PDFs, mean values, and quartile distributions of relative disk-averaged velocity $U_{\rm d}$ across several reference wind speeds $U_{\rm ref}$. Two wavelengths are simulated for the intermediate  $U_{\rm ref}=12$ m/s case. Left (blue) sides correspond to fixed bottom, while right (red) sides to floating (opposite phase) cases. Thus, from left to right, cases correspond to FB-10, OP-10, FB-12, OP-12, FB-SW-12, OP-SW-12, FB-15, OP-15, FB-18, and OP-18.} 
    \label{fig: V Violin}
\end{figure}

\begin{figure}[h!]
    \centering
    \includegraphics[width=\columnwidth]{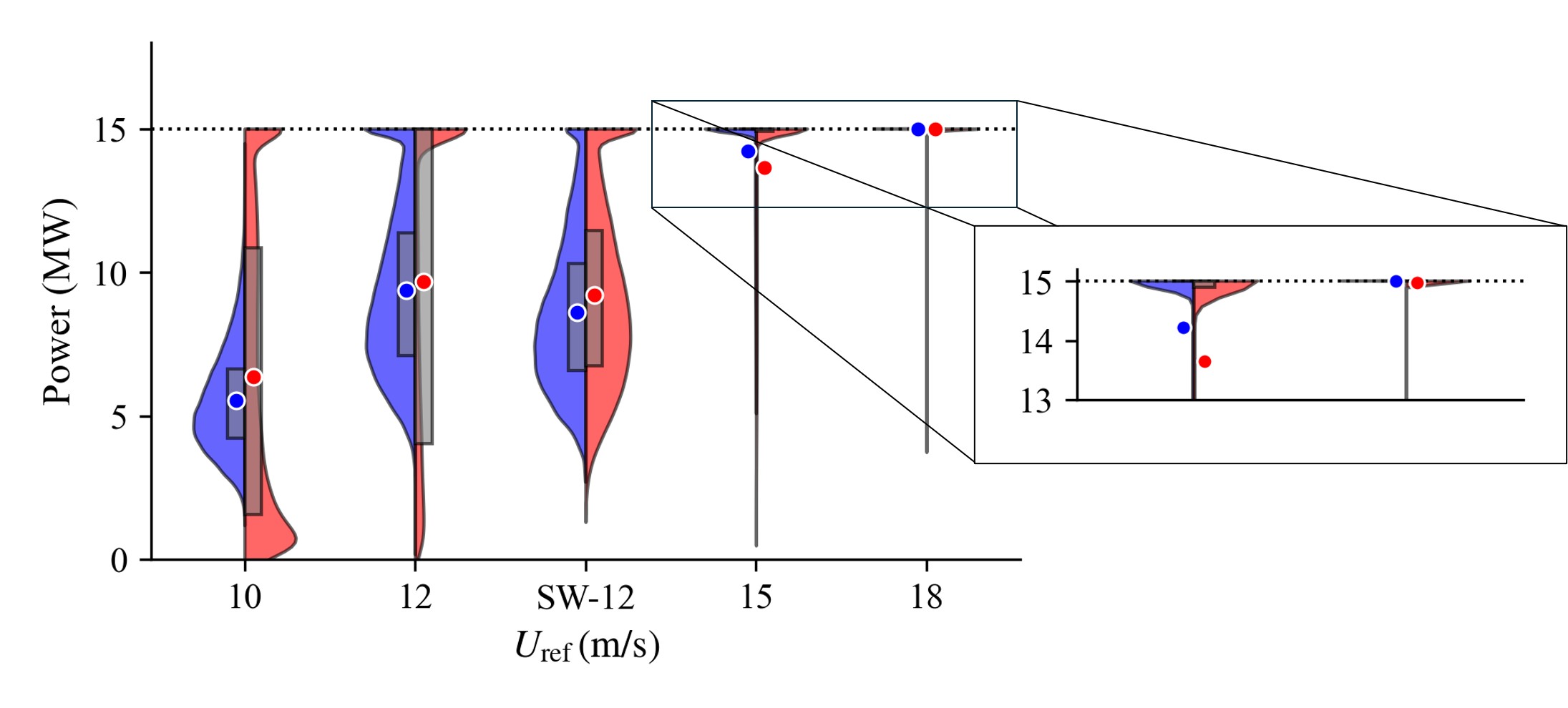}
    \caption{Same as Fig. \ref{fig: V Violin} but for turbine power. Violin plots showing PDFs, mean values, and quartile distributions of turbine power across several reference wind speeds $U_{\rm ref}$. Two wavelengths are simulated for the intermediate  $U_{\rm ref}=12$ m/s case. Left (blue) sides correspond to fixed bottom, while right (red) sides to floating (opposite phase) cases. Thus, from left to right, cases correspond to FB-10, OP-10, FB-12, OP-12, FB-SW-12, OP-SW-12, FB-15, OP-15, FB-18, and OP-18.}
    \label{fig: P Violin}
\end{figure}

Figs.~\ref{fig: V Violin} and ~\ref{fig: P Violin} present violin plots of the relative disk-averaged velocity $U_{\rm d}^{\rm rel}$, and power output, respectively, summarizing the mean, PDF, and interquartile ranges across cases with different reference wind forcing velocities. The cases are ordered along the $x-$axis with increasing $U_{\rm ref}$. The case "SW-12" denotes the $\lambda = S_x/15.5$ and $U_{\rm ref} = 12~\mathrm{m/s}$ case, while the remaining cases correspond to $\lambda = S_x/8.5$ with different values of $U_{\rm ref}$. In each violin plot, the left (blue) side represents FB turbines while the right (red) side corresponds to the opposite phase (OP) floating  turbines.

In Fig.~\ref{fig: V Violin}, the dotted horizontal lines denote the transition velocities between operational Regions 1 to 1.5, 1.5 to 2, and 2 to 3 (from bottom to top). Most turbines operate across multiple regions, with several OP cases (e.g., OP-10 and OP-12) spanning all four regions, indicating that transitions between operational regions are common. Under PCADM, the piecewise dependence of the turbine aerodynamic response on $U_{\rm d}^{\rm ref}$ introduces changes in the rotor-disk feedback as the turbine transitions between operating regions. In particular, when $U_{\rm d}^{\rm ref}$ crosses the transition velocity $U_{\rm d, tr}$, the thrust coefficient changes from independent of $U_{\rm d}$ to a decreasing function of $U_{\rm d}$, modifying the local thrust force and subsequently influencing the probability distribution of $U_{\rm d}$. Among the FB cases, only FB-15 exhibits a noticeable narrow peak near $U_{\rm d,tr}$. This occurs because the turbine frequently passes through the Region 2–3 transition under this condition, repeatedly experiencing changes in aerodynamic feedback. In contrast, the other FB cases remain predominantly within either Region 2 or Region 3 and therefore $U_{\rm d}$ exhibits approximately Gaussian-like distributions.

For OP cases, platform motion introduces an additional periodic velocity fluctuation that is superposed on the turbulent inflow variability, generally producing bimodal $U_{\rm d}^{\rm ref}$ distributions. The two peaks correspond to different phases of the turbine motion, where the rotor experiences distinct ranges of disk velocity. The only exception is the short-wave OP case, where the shorter wave period and smaller pitching amplitude reduce the separation between the two velocity states, causing the two peaks to merge into a single broader distribution.

As shown in Fig.~\ref{fig: P Violin}, PCADM appropriately limits the turbine power output to the rated value, resulting in saturation near the rated power level. Similar to the $U_{\rm d}$ distributions, OP turbines generally exhibit larger power fluctuations than FB turbines due to the additional unsteady effects introduced by platform motion, while the FB and OP cases under the short-wave condition (SW-12) show comparable power variability. Comparisons among cases with the  wavelength but various reference wind velocities show that 
the effect of turbine pitching on power depends strongly on the operational region. At low wind speeds (e.g., $U_\text{\rm ref} = 10,\mathrm{m/s}$, Region 2-dominated), floating turbines with larger velocity oscillations produce higher power due to the nonlinear cubic dependence of power on velocity. At higher wind speeds (e.g., $U_\text{\rm ref} = 15,\mathrm{m/s}$), the operation shifts mainly to Region 3 where power is constrained to the rated value. Here, low-velocity excursions due to periodic turbine pitching motions reduce output without compensating gains from high-velocity excursions. At intermediate wind speeds, competing effects lead to either higher or lower floating turbine power relative to fixed-bottom turbines, depending on the distribution of time spent in each operational region. When both turbine types operate entirely in Region 3, no discernable difference in power is observed. 

In addition to wind forcing, wave conditions further modulate turbine performance. Fig.~\ref{fig: V Violin} shows that under identical wind forcing of $U_{\rm ref}=12$, longer and faster waves generally promote higher power generation. Both FB-12 and FB-SW-12 remain largely in Region 2, while clear differences emerge between OP-12 and OP-SW-12 due to wave-induced platform motion. Case OP-SW-12 produces smaller pitching amplitudes, maintaining the turbine predominantly in Region 2, where it operates near the optimal $C_p$. Since both OP-SW-12 and FB-SW-12 spend comparable amounts of time in Region~2, the larger velocity oscillations in the floating case lead to enhanced power production. Conversely, case OP-12 generates larger motions that spread operation across multiple regions. Nearly half of the operating time occurs in Region~3, where the turbine produces near-rated power, whereas FB-12 remains predominantly in Region~2. This shift in operational regime explains the more pronounced power enhancement observed in OP relative to FB under short-wave conditions compared with long-wave conditions. 

Overall, these results demonstrate that turbine pitching effects on power are intrinsically dependent on the operational region and are strongly modulated by combined wind–wave conditions. Further they show that PCADM instead of CCADM is required for physically consistent predictions of turbine performance when modeling floating turbines in LES using ADM. 
\subsection{\label{sec:Phase} Effects of Turbine-Wave Phase}

In this section the effect of pitching phase on relative velocity and turbine power is investigated by examining pitching motions with different wave-induced turbine motion phase offsets. The OP and FB turbines share the same wave configuration, while for the IP and IC cases the wavelengths are adjusted slightly to create distinct turbine motion phase differences while ensuring the same turbine spacings. 

\begin{figure}
    \centering
    \includegraphics[width=1.0\columnwidth]{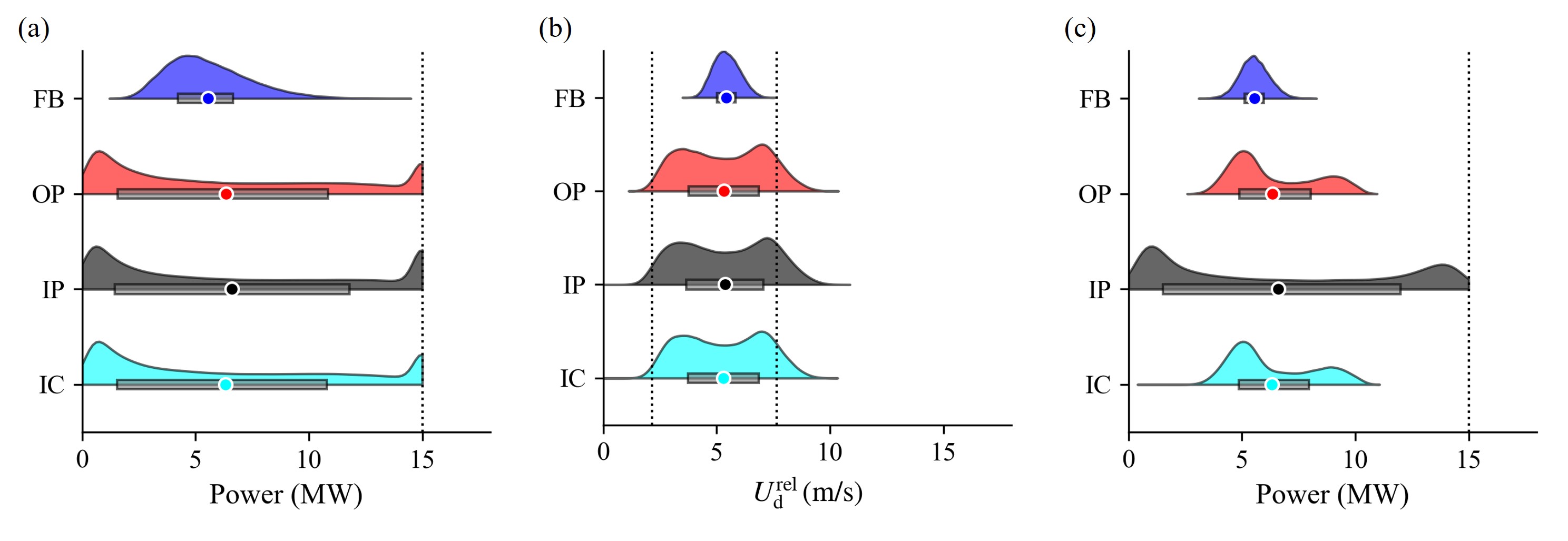}    
    \label{fig: ADM Comp.}
    \caption{PDFs of individual turbine velocities (a), and power (b), and turbine farm scale average turbine power (c, average over 6 turbines) comparing FB, OP, IP and IC pitching motions as indicated, for the reference wind speed of $U_{\rm ref} = 10 \mathrm{m/s}$. From top to bottom, cases correspond to FB-10, OP-10, IP-10, and IC-10.}
  \label{fig: Phases_10}
\end{figure}

Fig.~\ref{fig: Phases_10}  shows PDFs of power and relative disk-averaged velocity $U_{\rm d}^{\rm rel}$ for different turbine motion phases (IP, OP and IC), compared with FB turbine at the reference speed of $U_{\rm ref}=10\mathrm{m/s}$, i.e., cases IP-10, OP-10, IC-10 and FB-10. Panels (a) and (b) correspond to results for representative single turbines, while panel (c) shows the wind farm average power over 6 turbines. Consistent with prior observations, all floating turbines exhibit larger oscillations compared to the FB case. As expected, average power generation variability at the farm level is generally reduced relative to the turbine level due to averaging across 6 turbines.  

\begin{figure}
    \centering
    \includegraphics[width=1.0\columnwidth]{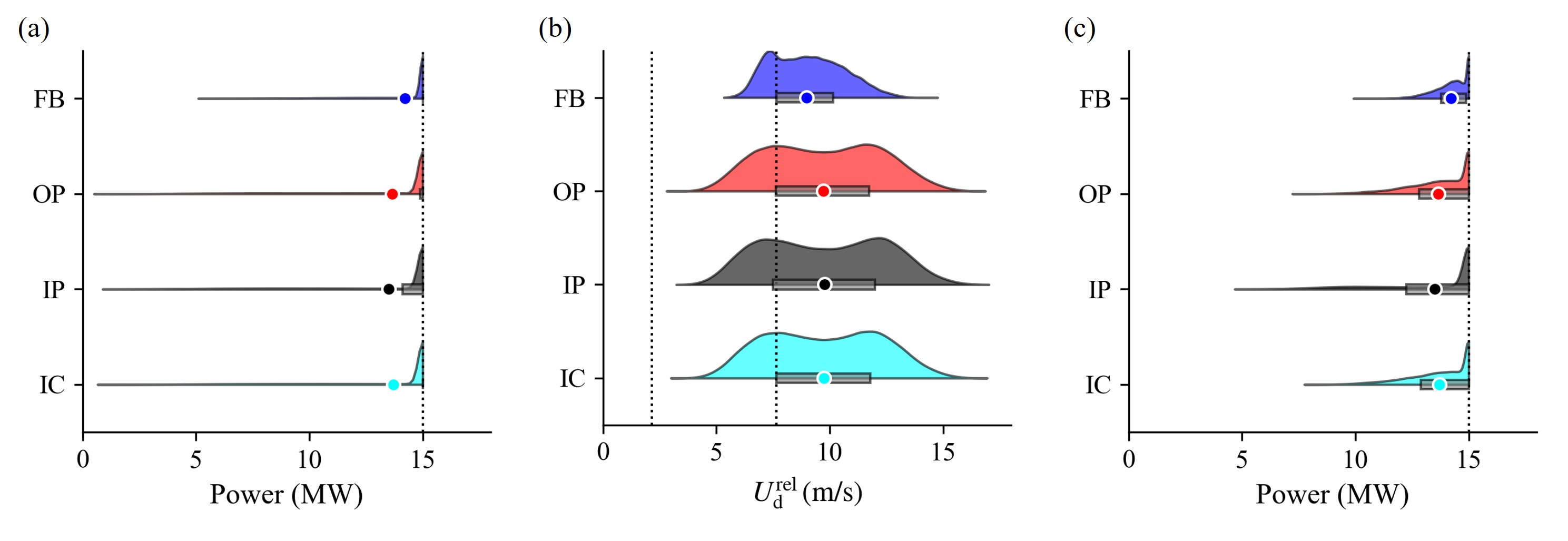}    
    \label{fig: ADM Comp.}
    \caption{Same as Fig. \ref{fig: Phases_15} but for the wind speed $U_{\rm ref} = 15\mathrm{m/s}$. PDFs of individual turbine velocities (a), and power (b), and turbine farm scale average turbine power (c, average over 6 turbines) comparing FB, OP, IP and IC pitching motions as indicated, for the reference wind speed of $U_{\rm ref} = 15 \mathrm{m/s}$. From top to bottom, cases correspond to FB-15, OP-15, IP-15, and IC-15.}
  \label{fig: Phases_15}
\end{figure}

As the reference wind velocity is increased  to $U_{\rm ref}=15\mathrm{m/s}$ (Fig.~\ref{fig: Phases_15}), turbine operation transitions from Region 2-dominated to Region 3-dominated behavior. Unlike the $U_{\rm ref}=10\mathrm{m/s}$ case, the power distribution of $U_{\rm ref}=15\mathrm{m/s}$ cases shows saturation at rated power value due to the plateau behavior of the power curve in Region 3. Two distinct peaks associated with the harmonic motion of floating turbines are evident in the $U_{\rm d}^{\rm rel}$ distribution for both the $U_{\rm ref}=10\mathrm{m/s}$ and $15\mathrm{m/s}$ cases.

At the turbine level, no significant differences are observed among the floating cases with different phases. In contrast, at the farm level, the IP configuration exhibits much larger power variability, consistent with the inter-turbine correlations, whereas the variability in the OP and IC cases is strongly reduced due to cancellations of positive and negative deviations. While the extrema and interquartile ranges shift between turbine and farm level analyses, the mean values are the same, as expected from linear averaging.

\subsection{\label{sec: wave and zo_lo}
Quantifying the effects of wave motion parameterization on the results
}

As discussed previously, the wave field alters the ocean’s aerodynamic roughness and the air–sea momentum flux through the relative velocity between the incoming wind and the moving wave surface. These effects feed back into the airflow, wind speed and turbulence profile in the boundary layer. The MOSD framework estimates the net pressure force arising from the velocity difference between the wind speed and wave phase speed.
Because the model is formulated directly from the local surface slope and its phase speed, wave-phase information is explicitly incorporated into the modeling of air–water momentum transfer. By comparing results using the phase-dependent MOSD wall model with the phase-averaged EQWM formulation across both FB and floating turbines (OP), we can assess how wave-phase aware wave modeling affects the 
statistics of the generated turbine power. However, for a meaningful comparison, an appropriate value of the roughness length $z_0$ used in the EQWM must be selected. It is obtained here from separate  MOSD simulations conducted with identical domain and turbine configurations. The value of $z_0$ is determined by fitting the horizontally averaged mean velocity profile below the turbine to the logarithmic law of the wall using a least-squares approach. The friction velocity $u^*$ is taken as the time- and spatially averaged value of the surface stress extracted from the LES results.  For the EQWM-based simulations, the turbine motion is prescribed through an imposed $\theta(t)$ in the same manner as the cases using MOSD, ensuring that both the frequency and amplitude of the turbine motion match those in the MOSD simulations. 

\begin{figure}[htbp!]
    \centering
    \includegraphics[width=0.8\columnwidth]{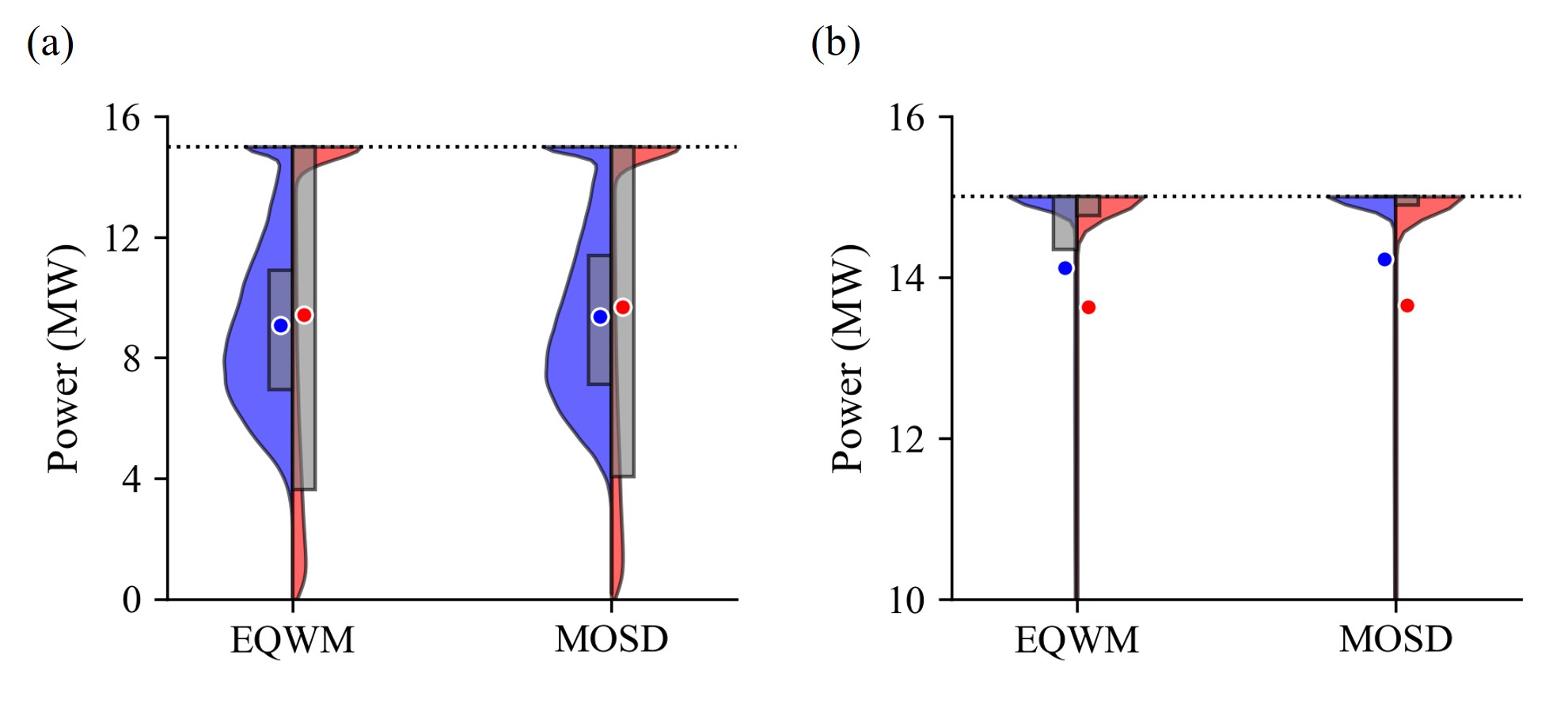}    
    \caption{Violin plots of turbine power comparing EQWM with MOSD wall model under (a) $U_{\rm ref} = 12 \mathrm{m/s}$ cases (FB-EQ-12, OP-EQ-12, FB-12, and OP-12) and (b) $U_{\rm ref} = 15 \mathrm{m/s}$ cases (FB-EQ-15, OP-EQ-15, FB-15, and OP-15). Left (blue) sides correspond to fixed bottom, while right (red) sides to floating (opposite phase) cases.}

    \label{fig: Wave}
\end{figure}

The violin plots shown in Fig.~\ref{fig: Wave} present the PDF, mean and quartile values of predicted power  FB (blue) and OP (red) cases at both $U_{\rm ref}=12\mathrm{m/s}$ (FB-EQ-12, OP-EQ-12, FB-12, and OP-12) in panel (a) and $U_{\rm ref}=15\mathrm{m/s}$ (FB-EQ-15, OP-EQ-15, FB-15, and OP-15) in panel (b), using both MOSD and EQWM wall models. 
Consistent with the MOSD results, the EQWM framework predicts broader power distributions for OP compared to FB. The mean values of predicted power are essentially unchanged when comparing MOSD to EQWM results, suggesting that the dominant differences observed between FB and OP turbines are governed by wave-induced relative velocities between the wind and the turbine rather that phase-dependent wave effects on the wind flow, which may affect the turbine response. However, the results depend significantly on the imposed value of $z_0$ whose determination required the use of MOSD.

\begin{table*}
\caption{\label{tab:results} Key flow and turbine parameters determined from LES for different cases.}
\label{tb: Results}
\begin{ruledtabular}
\begin{threeparttable}
\begin{tabular}{c c c c c c c c}
Case
& $\overline{U}_{\rm d}^{\rm rel}$ (m/s)
& $\overline{P}$ (MW)
& \% in Region~2 \\
\hline
FB-15
& 9.0
& 14.2
& 25.0 \\

FB-C-15
& 8.1 
& 18.0
& -- \\

OP-15
& 9.7
& 13.7
& 25.2 \\

OP-C-15
& 7.9 
& 20.8
& -- \\

IP-15 
& 9.8
& 13.5
& 26.9 \\

IC-15
& 9.7
& 13.7
& 24.9 \\

FB-EQ-15
& 8.8
& 14.1
& 27.9\\

OP-EQ-15
& 9.7
& 13.6
& 25.5\\

FB-10
& 5.4
& 5.6
& 100.0\\

OP-10
& 5.4
& 6.4
& 89.9 \tnote{a}\\

IP-10
& 5.4
& 6.7
& 85.9 \tnote{b} \\

IC-10
& 5.3
& 6.4
& 90.0 \tnote{c} \\

FB-12
& 6.4
& 9.4
& 92.7 \\

OP-12
& 7.0
& 9.7
& 57.1 \tnote{d} \\

FB-EQ-12
& 6.4
& 9.1
& 95.4\\

OP-EQ-12
& 6.9
& 9.4
& 58.7 \tnote{e}\\

FB-18 
& 12.8
& 15.0
& 0.0\\

OP-18
& 12.7
& 15.0
& 0.8 \\

FB-SW-12
& 6.3
& 8.6
& 97.2\\

OP-SW-12
& 6.4
& 9.2
& 92.0\\
\end{tabular}
\begin{tablenotes}
\item[a-e] Region~2 percentages include a small contribution from Region~1: $a = 1.2\%$, $b = 2.6\%$, $c = 1.5\%$, $d = 0.3\%$,  and $e = 0.4\%$.
\end{tablenotes}
\end{threeparttable}
\end{ruledtabular}
\end{table*}

\section{\label{sec:Conc}Conclusions}

In this study, we investigated the power generated by   floating and fixed-bottom wind turbines under a range of wind–wave conditions using LES. 
We first demonstrated that LES employing the traditional CCADM (constant $C_T$ and $C_p$) which has been often adopted in laboratory and numerical studies, can lead to unphysical results such as overpredictions of power output for floating turbines operating under high wind speeds and strong wave conditions. Overpredictions are typically due to periodic relative velocity excursions above the turbine's rated speed. To address this limitation, we employed a more realistic representation through the PCADM. It accounts for dependence of $C_T$ and $C_p$ on relative inflow velocity differing in different turbine operational regions, thus properly limiting power to its rated value and reflecting additional operational trends. Comparisons between CCADM and PCADM 
demonstrated that incorporating realistic $C_T$ and $C_P$ behavior is essential for predicting meaningful offshore wind turbine performance in LES.

Using the PCADM framework, we further examined turbine responses under various wind–wave conditions. The results show that deciding whether floating and fixed-bottom turbines generate more or less power depends strongly on the turbine operating region, which is determined by the coupled effects of wind forcing and wave-induced harmonic platform motion. Under Region 2 dominant conditions, floating turbines can yield enhanced power output due to wave-induced cyclic relative velocity variations and the nonlinear cubic dependence of power on relative disk normal velocity. In contrast, when operation is predominantly in Region 3, power regulation limits the output, and motion-induced low-velocity excursions reduce floating turbine performance without compensating gains, leading to slightly lower or comparable power generation compared to fixed-bottom turbines. At intermediate wind speeds, the net power difference depends on the fraction of time spent across operational regions. When turbines are limited to their rated power production, higher mean relative disk-normal velocities do not necessarily translate into higher overall power output.

We also examined the influence of different relative phases between turbines that can occur due to different ratios of inter-turbine spacings and incoming ocean wave lengths (in phase, out of phase, or inconmensurate phase). At the single-turbine level, both the relative velocity and corresponding power output exhibit very little sensitivity to  phase. However, when averaging across turbines (relevant for farm-level power outputs) the IP configuration produces larger fluctuations, caused by stronger turbine–turbine correlations compared to the OP and IC cases, which could have implications for turbine loading. Finally, we demonstrated that LES employing the traditional wave phase averaged EQWM yielded negligible differences in mean power $\overline{P}$, but the corresponding effective roughness height above the turbines, $z_0$, cannot be determined a priori without performing MOSD or wave-resolving simulations and fitting the upper logarithmic velocity profile. This implies that MOSD remains a practical approach for offshore wind LES simulations for which $z_0$ would not be known. Overall, these findings highlight the importance of realistic representations of turbine operating region and platform-induced variability in LES frameworks to achieve accurate predictions of turbine performance under coupled wind–wave conditions.

Finally, we note several limitations of the present study that should be considered when interpreting the results. We only used simple, regular waves, and the analysis was restricted to a limited set of wave conditions, primarily involving two representative wavelengths and monochromatic wave configurations. In addition, platform motion was prescribed in a simplified form and did not include the full range of coupled surge, heave, and multi-directional wave-induced responses. While this framework enables the systematic investigation of idealized turbine pitch motion and its impact on power generation under various wind–wave conditions in a fully developed wind turbine array marine boundary layer, it does not capture turbine responses under more realistic sea states. For instance, nonlinear interactions between multiple degrees of freedom and turbine aerodynamics were not fully captured. These limitations suggest that the present results should be interpreted as a baseline characterization of wind–wave–turbine coupling under idealized conditions. Future work should extend this framework to irregular wave spectra, fully coupled multi-DOF platform dynamics, and a wider range of farm layouts and operating conditions to establish more generalizable predictions of offshore wind turbine and wind farm performance.

\section{Acknowledgment}
This work was partially supported by the US National Science Foundation (grant number CBET-2401013) and the Academic Center For Reliability \& Resilience of Offshore Wind (ARROW), which was initially supported by the U.S. Department of Energy’s Office of Energy Efficiency and Renewable Energy (WETO DE-EE0011269). Continuing ARROW support from the Massachusetts Clean Energy Center and the Maryland Energy Administration is gratefully acknowledged. Computations were performed at the Advanced Research Computing at Hopkins (ARCH) core facility (rockfish.jhu.edu), that was initiated through by the US NSF grant number OAC1920103.

\appendix

\section{\label{sec:other turbines}PCADM fitted $C_T$ and $C_p$ curves for two other reference wind turbines}

In this study, the IEA~15 MW reference turbine is used to demonstrate the proposed procedure, but the PCADM framework is readily applicable to other turbine systems. To illustrate this generality, the $C_T$ and $C_p$ characteristics of the NREL~5MW and IEA~22MW turbines are also considered. For the NREL 5 MW turbine \cite{Jonkman_et_al_2009}, Region 1.5 is explicitly distinguished in both $C_T$ and $C_p$. The $C_T$ relation is modified by introducing a Region 1.5 branch, as given in Eq.~\ref{eq: CT-5MW}, with $C_{\rm T,op}=0.75$, $\alpha_1=0.45$, and $\alpha_2=6.84$. 
\begin{equation}
C_T(U_\infty) =
\begin{cases}
0, &  U_\infty\leq U_{ci}\\
\dfrac{C_{\rm T,op}}{1+\alpha_1(1 - U_\infty/U_{\rm tr,1})}, & U_{\rm ci} < U_\infty \leq U_{\rm tr, 1}\\
C_{\rm T,op}, & U_{\rm tr, 1}<U_\infty \leq U_{\rm tr}\\[3pt]
\dfrac{C_{\rm T,op}}{1+\alpha_2(U_\infty/U_{\rm tr} - 1)}, & U_{\rm tr} < U_\infty \leq U_{\rm co}\\
0, &  U_{\rm co} < U_\infty\\
\end{cases}
\label{eq: CT-5MW}
\end{equation}
Following the same procedure from Eqs.~\ref{eq: a-U_inf} -~\ref{eq: a-U_d}. The $a-U_{\rm d}$ relation can be written as 
\begin{equation}
a(U_{\rm d}, \theta) = \dfrac{1}{2}
\begin{cases}
0, &  U_{\rm d} \leq U_{\rm d,tr1}\\
1+\dfrac{\alpha_1}{\alpha_1+1}\dfrac{U_{\rm d}}{U_{\rm tr1}}-\sqrt{\left( 1+\dfrac{\alpha_1}{\alpha_1+1}\dfrac{U_{\rm d}}{U_{\rm tr1}}\right)^2+\dfrac{C_{\rm T,op}\cos{\theta}}{\alpha_1+1}}, & U_{\rm d,tr1} < U_{\rm d} \leq U_{\rm d,ci}\\[5pt]
1-\sqrt{1-C_{\rm T,op}\cos{\theta}}, & U_{\rm ci}<U_{\rm d} \leq U_{\rm tr}\\[5pt]
1-\dfrac{\alpha_2}{\alpha_2-1}\dfrac{U_{\rm d}}{U_{\rm tr}}+\sqrt{\left( 1-\dfrac{\alpha_2}{\alpha_2-1}\dfrac{U_{\rm d}}{U_{\rm tr}}\right)^2+\dfrac{C_{\rm T,op}\cos{\theta}}{\alpha_2-1}}, & U_{\rm d,tr} < U_{\rm d} \leq U_{\rm d,co}\\
0, &   U_{\rm d,co} < U_{\rm d}.
\end{cases}
\label{eq: a-5MW}
\end{equation}

For NREL~5MW turbine, the $C_p$ formulation (Eq.~\ref{eq: CP-U_inf}) remains unchanged, with parameters $P_{\rm rt}=5~\mathrm{MW}$, $C_p=0.482$, $U_{\rm ci}=3~\mathrm{m/s}$, $U_{\rm tr1}=7.8~\mathrm{m/s}$, $U_{\rm tr}=11.4~\mathrm{m/s}$, $U_{\rm co}=25~\mathrm{m/s}$, $\gamma=0.45$, $\sigma=3.7$. $U_{\rm d,ci} =  \dfrac{U_{\rm ci}}{2} \, \left(1+\sqrt{1-C_{\rm T}\cos{\theta}}\right)$; $U_{\rm d,tr1}=
\dfrac{U_{\rm tr1}}{2} \,
\left(
1+\sqrt{1-C_{\rm T}\cos\theta}
\right)$; $U_{\rm d,tr}=
\dfrac{U_{\rm tr}}{2} \,
\left(
1+\sqrt{1-C_{\rm T}\cos\theta}
\right)$, and $U_{\rm d,co} =  \dfrac{U_{\rm co}}{2} \, \left(1+\sqrt{1-C_{\rm T}\cos{\theta}}\right)$.

\begin{figure}[h!]
    \centering
    \includegraphics[width=0.9\columnwidth]{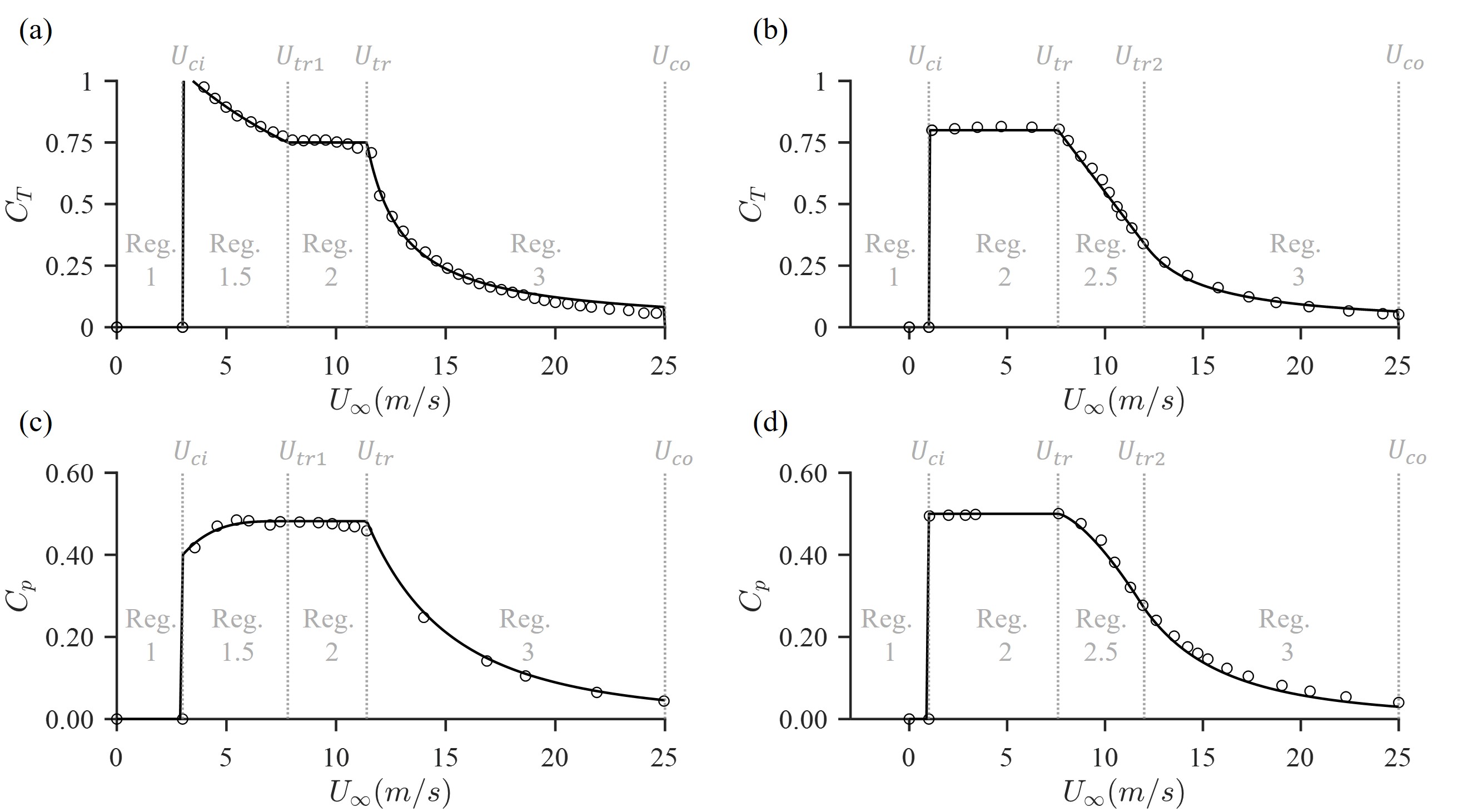}    
    \caption{Thrust coefficient $C_T$ and power coefficient $C_p$ as functions of wind speed for two turbines. Panels (a,b) correspond to the NREL~5MW turbine, and panels (c,d) to the IEA~22MW turbine. In each case, $C_T$ is shown in (a,c) and $C_p$ in (b,d). Solid lines denote PCADM predictions and circular markers indicate reference controller data.}
  \label{fig: CTCP}
\end{figure}

For the IEA~22MW turbine, Regions 1.5 is neglected and Region~2.5 occurs. The $C_T$ and $C_p$ relations are modified by introducing a Region 2.5 branch. The piecewise function for $C_T$ is fitted as
\begin{equation}
C_T(U_\infty) =
\begin{cases}
0, &  U_\infty\leq U_{\rm ci}\\
C_{\rm T,op}, & U_{\rm ci} < U_\infty \leq U_{\rm tr, 2}\\[3pt]
C_{\rm T,op} - \alpha_3(U_\infty/U_{\rm tr2}-1);~~~~~~\alpha_3 =  \dfrac{C_{\rm T,op} - C_{\rm T,tr}} {U_{\rm tr}/U_{\rm tr2}-1}, & U_{\rm tr, 2} < U_\infty \leq U_{\rm tr}\\
\dfrac{C_{\rm T,tr}}{1+\alpha_4(U_\infty/U_{\rm tr} - 1)}, & U_{\rm tr} < U_\infty \leq U_{\rm co}\\
0, &  U_{\rm co} < U_\infty.\\
\end{cases}
\label{eq: CT-22MW}
\end{equation}
For this turbine, the fitted parameters are $C_{\rm T,op}=0.5$, $C_{\rm T,tr} = 0.34$, $U_{\rm ci}=1~\mathrm{m/s}$, $U_{\rm tr2}=7.6~\mathrm{m/s}$, $U_{\rm tr}=12.0~\mathrm{m/s}$, $U_{\rm co}=25~\mathrm{m/s}$, $\alpha_4=4$. Similarly, the $a-U_{\rm d}$ relation can be derived as
\begin{equation}
a(U_{\rm d}, \theta) = \dfrac{1}{2}
\begin{cases}
0, &  U_{\rm d} \leq U_{\rm d,ci}\\
1-\sqrt{1-C_{\rm T,op}\cos{\theta}}, & U_{\rm d,ci}<U_{\rm d} \leq U_{\rm d,tr2}\\[5pt]
\frac{2}{3} + \sqrt[3]{-\frac{q}{2}+\sqrt{\Delta}} + \sqrt[3]{-\frac{q}{2}-\sqrt{\Delta}}, & U_{\rm d,tr2} < U_{\rm d} \leq U_{\rm d,tr}\\[5pt]
1-\dfrac{\alpha_4}{\alpha_4-1}\dfrac{U_{\rm d}}{U_{\rm tr}}+\sqrt{\left( 1-\dfrac{\alpha_4}{\alpha_4-1}\dfrac{U_{\rm d}}{U_{\rm tr}}\right)^2+\dfrac{C_{\rm T,op}\cos{\theta}}{\alpha_4-1}}, & U_{\rm d,tr} < U_{\rm d} \leq U_{\rm d,co}\\
0, &   U_{\rm d,co} < U_{\rm d}.
\end{cases}
\label{eq: a-22MW}
\end{equation}
Where
\begin{equation}
\begin{gathered}    
p
=
\frac{(C_{\rm T,op}+\alpha_3)\cos{\theta}-4}
{4}
-\frac{1}{3},~~~
q
=
\frac{2}{27}
+\frac{\cos{\theta}}{12}\left(
{\frac{3\alpha_3U_{\rm d}}{U_{\rm tr2}}-C_{\rm T,op}-\alpha_3} \right),~~~
\Delta
=
\left(\frac{q}{2}\right)^2
+\left(\frac{p}{3}\right)^3,
\\
U_{\rm d,ci}=
\dfrac{U_{\rm ci}}{2} \,
\left(
1+\sqrt{1-C_{\rm T}\cos\theta}
\right); \, U_{\rm d,tr2} = \dfrac{U_{\rm tr2}}{2}\left(1+\sqrt{1-C_{\rm T,op}\cos{\theta}}\right)
\\
U_{\rm d,tr}=
\dfrac{U_{\rm tr}}{2} \,
\left(
1+\sqrt{1-C_{\rm T}\cos\theta}
\right); \, U_{\rm d,co} = \dfrac{U_{\rm co}}{2}\left(1+\sqrt{1-C_{\rm T,op}\cos{\theta}}\right)
\end{gathered}
\end{equation}
The $C_P$ piecewise function for IEA~22MW turbine is
\begin{equation}
C_P(U_\infty) =
\begin{cases}
0, &  U_\infty \leq U_{ci}\\
C_{\rm P,op}, & U_{\rm ci}<U_\infty \leq U_{tr, 2}\\[3pt]
C_{\rm P, op}-\gamma_1(U_\infty/U_{\rm tr2}-1)^{\sigma_1} ;~~~~~~\gamma_1=\dfrac{C_{\rm P,op} - C_{\rm P,tr}} {(U_{\rm tr}/U_{\rm tr2}-1)^{\sigma_1}} & U_{tr, 2}<U_\infty \leq U_{\rm tr}\\[3pt]
C_{\rm P,tr}(\dfrac{U_{\rm tr}}{U_\infty})^3, &U_{\rm tr} < U_\infty \leq U_{\rm co}\\
0, &  U_{co} < U_\infty\\
\end{cases}
\label{eq: CP-22WM}
\end{equation}
The parameters are $C_{\rm p,op}=0.5$. $C_{\rm p,tr3} = C_p(U_\infty^{\rm rel}=U_{\rm tr3})=0.27$, $\sigma_1=1.45$. The resulting PCADM curves, fitted using Eqs.~\ref{eq: CT-5MW} and \ref{eq: CP-U_inf} for the NREL 5-MW turbine and Eqs.~\ref{eq: CT-22MW} and \ref{eq: CP-22WM} for the IEA 22-MW turbine, agree well with the manufacturer-provided data, as shown in Fig.~\ref{fig: CTCP}.

These fitted $C_T$ and $C_p$ curves enable straightforward application of the PCADM framework to both turbines. More generally, given manufacturer-provided $C_T$ and $C_p$ data, PCADM can be constructed for arbitrary turbine systems, providing a more realistic representation than CCADM for both FBWTs and OFWTs configurations.

\section{\label{sec:convergence}Statistical Convergence Error Analysis}

While LES is expected to yield statistically converged solutions given sufficiently long averaging times, single realizations run over finite time durations can exhibit errors due to insufficient statistical convergence. In particular, the mean values reported in this paper often  differ by small amounts and it is important to confirm that the differences observed (eg of mean power production between different cases) are larger than statistical convergence errors. Additionally, individual runs obtained under identical configurations and run for long times may still differ due to possible metastable states or lack of complete ergodicity that may cause differences among different runs. 
In this section, the sensitivity of reported mean values to averaging time and different realizations is quantified by  comparing running averages (Eq.~\ref{eq: Average}) obtained from three independent trials performed under identical LES setups, for selected representative cases. The running average of a variable $x(t)$ (such as turbine power or disk-averaged velocity) is computed according to
\begin{equation}
\overline{x}(t_n) = \frac{1}{n} \sum_{i=1}^{n} x(t_i)
\label{eq: Average}
\end{equation}
where 
$n$ is the number of consecutive time samples used in the averaging. In all cases, the number of samples is varied between 1 and reaching up to $n_{\rm max} = 1{,}500{,}000$. The LES time step is variable, determined by a fixed CFL number ($CFL = 0.0625$). The maximum total physical averaging times therefore varies slightly across different cases but typically reaching approximately between $t_{n_{\rm max}} = 8\times ^4$ and $10^5$ seconds (around 1 day). The running averages are computed after a sufficiently long initial transient period to ensure that the statistics are representative of a converged state. 

\begin{figure}
    \centering
    \includegraphics[width=0.9\columnwidth]{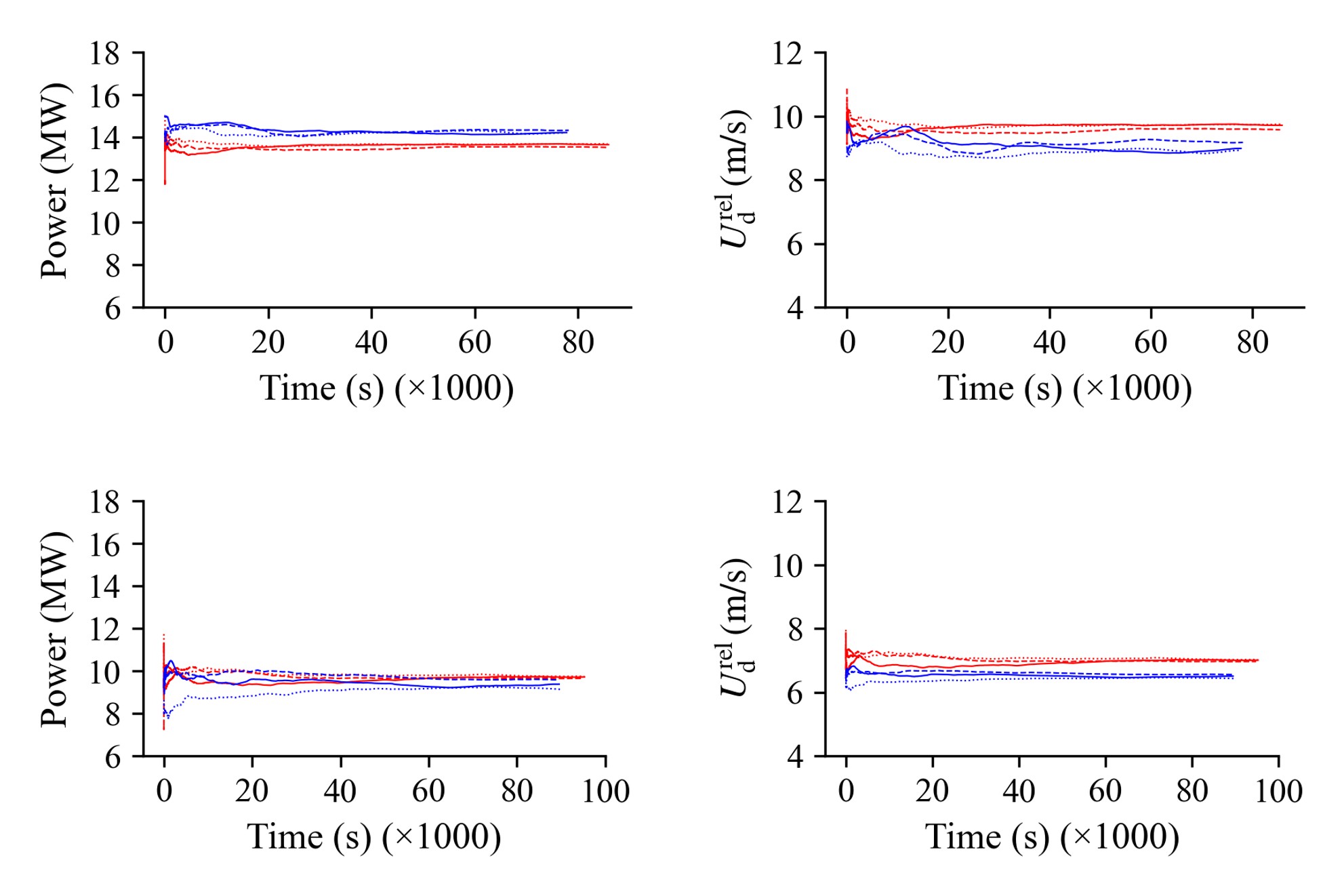}    
    \caption{Running averages from three independent LES trials for FB (blue lines) and OP (red lines) configurations. Power is shown in the left column and $U^{\rm rel}_{\rm d}$ in the right column. The top row corresponds to $U_{\mathrm{ref}} = 15~\mathrm{m/s}$ and the bottom row to $U_{\mathrm{ref}} = 12~\mathrm{m/s}$.}
    \label{fig: Converge}
\end{figure}

Figure~\ref{fig: Converge} presents the running averages of power and relative disk-normal velocity $U_\infty$ for the FB and OP configurations under two incoming flow conditions. The top row corresponds to $U_{\mathrm{ref}} = 15~\mathrm{m/s}$, and the bottom row corresponds to $U_{\mathrm{ref}} = 12~\mathrm{m/s}$. As shown in Fig.~\ref{fig: Converge}, the running averages exhibit noticeable fluctuations during the early phase but gradually stabilize, indicating statistical convergence of the LES results. The remaining differences between trials are relatively small. For power, the observed trial-to-trial variations are $\Delta P = 0.171$ MW for case OP-15 (i.e. 1.10\% of $P_{rt}$), $\Delta P = 0.120$ MW for case FB-15 (0.79\% of $P_{rt}$), $\Delta P = 0.443$ MW for case IP-12 (2.95\% of $P_{rt}$), and $\Delta P  = 0.030$ MW for case FB-12 (0.20\% of $P_{rt}$). For the relative disk-normal velocity, the corresponding variations are $\Delta U_\infty = 0.143$ m/s  for case OP-15 (1.35\% of $U_{\rm tr}$), $\Delta U_\infty = 0.155$ m/s for case FB-15 (1.46\% of $U_{tr}$), $\Delta U_\infty = 0.153$ (1.44\% of $U_{\rm tr}$) for case OP-12, and $\Delta U_\infty = 0.054$ m/s for case FB-12 (0.05\% of $U_{\rm tr}$). Overall, the variability across independent LES trials is smaller than the differences observed between mean values in power reported in the main text, for instance between the OP and FB configurations, particularly visible for $U_{\mathrm{ref}} = 15~\mathrm{m/s}$, where both configurations operate in Region 2 for comparable durations. This analysis  therefore reinforces the statistical robustness of the reported trends. 

\newpage

\bibliography{aipsamp}

@article{Calaf_Meneveau_Meyers_2010,
  title   = {Large eddy simulation study of fully developed wind-turbine array boundary layers},
  author  = {Calaf, Marc and Meneveau, Charles and Meyers, Johan},
  journal = {Physics of Fluids},
  volume  = {22},
  number  = {1},
  pages   = {015110},
  year    = {2010},
  doi     = {10.1063/1.3291077}
}

@article{Shapiro_Gayme_Meneveau_2018, 
title={Modelling yawed wind turbine wakes: a lifting line approach}, 
volume={841}, 
DOI={10.1017/jfm.2018.75}, 
journal={Journal of Fluid Mechanics}, 
author={Shapiro, Carl R. and Gayme, Dennice F. and Meneveau, Charles}, 
year={2018}, 
pages={R1}}

@article{Heck_Johlas_Howland_2023, 
title={Modelling the induction, thrust and power of a yaw-misaligned actuator disk}, 
author={Heck, K. S. and Johlas, H. M. and Howland, M. F.}, 
journal={Journal of Fluid Mechanics}, 
year={2023}, 
volume={959}, 
pages={A9},  
DOI={10.1017/jfm.2023.12}
}

@techreport{gaertner2020iea15mw,
  author       = {Gaertner, Evan and Rinker, Jennifer and Sethuraman, Latha and Zahle, Frederik and Anderson, Benjamin and Barter, Garrett and Abbas, Nikhar and Meng, Fanzhong and Bortolotti, Pietro and Skrzypinski, Witold and Scott, George and Feil, Roland and Bredmose, Henrik and Dykes, Katherine and Shields, Matt and Allen, Christopher and Viselli, Anthony},
  title        = {Definition of the IEA 15-Megawatt Offshore Reference Wind Turbine},
  institution  = {National Renewable Energy Laboratory},
  address      = {Golden, CO},
  year         = {2020},
  number       = {NREL/TP-5000-75698},
  url          = {https://www.nrel.gov/docs/fy20osti/75698.pdf}
}

@book{Burton2021WindEnergyHandbook,
  title        = {Wind Energy Handbook},
  author       = {Burton, Tony and Jenkins, Nick and Bossanyi, Ervin and Sharpe, David and Graham, Michael},
  edition      = {3rd},
  year         = {2021},
  publisher    = {John Wiley \& Sons},
  address      = {Chichester, UK},
  isbn         = {9781119642331}
}

@article{Stevens_Meneveau_2017,
  title   = {Flow Structure and Turbulence in Wind Farms},
  author  = {Stevens, Richard J.A.M. and Meneveau, Charles},
  journal = {Annual Review of Fluid Mechanics},
  volume  = {49},
  number  = {1},
  pages   = {311--339},
  year    = {2017},
  doi     = {10.1146/annurev-fluid-010816-060206}
}

@article{Xiao_Yang_2019,
  title   = {Large-Eddy Simulation-Based Study of Effect of Swell-Induced Pitch Motion on Wake-Flow Statistics and Power Extraction of Offshore Wind Turbines},
  author  = {Xiao, Shuolin and Yang, Di},
  journal = {Energies},
  volume  = {12},
  number  = {7},
  pages   = {1246},
  year    = {2019},
  doi     = {10.3390/en12071246}
}

@article{Narasimhan_Gayme_Meneveau_2022,
  title   = {Effects of wind veer on a yawed wind turbine wake in atmospheric boundary layer flow},
  author  = {Narasimhan, Ghanesh and Gayme, Dennice F. and Meneveau, Charles},
  journal = {Physical Review Fluids},
  volume  = {7},
  number  = {11},
  pages   = {114609},
  year    = {2022},
  doi     = {10.1103/PhysRevFluids.7.114609}
}

@article{Xiao_Zhu_Narasimhan_Gayme_Meneveau_2025,
  title   = {Wind farm dynamics over a diurnal cycle: Analysis of a comprehensive large-eddy simulation, web-services accessible dataset},
  author  = {Xiao, Shuolin and Zhu, Xiaowei and Narasimhan, Ghanesh and Gayme, Dennice F. and Meneveau, Charles},
  journal = {Journal of Renewable and Sustainable Energy},
  volume  = {17},
  number  = {6},
  pages   = {063301},
  year    = {2025},
  doi     = {10.1063/5.0283412}
}

@article{Kasper_Stieren_Stevens_2024,
  title   = {Simulation and modeling of wind farms in baroclinic atmospheric boundary layers},
  author  = {Kasper, J. H. and Stieren, A. and Stevens, R. J. A. M.},
  journal = {Journal of Renewable and Sustainable Energy},
  volume  = {16},
  number  = {6},
  pages   = {063302},
  year    = {2024},
  doi     = {10.1063/5.0220322}
}

@techreport{Bak_et_al_2013,
  title       = {The DTU 10-MW Reference Wind Turbine},
  author      = {Bak, Christian and Zahle, Frederik and Bitsche, Robert and Kim, Taeseong and Yde, Anders and Henriksen, Lars Christian and Hansen, Morten Hartvig and Blasques, José Pedro Albergaria Amaral and Gaunaa, Mac and Natarajan, Anand},
  institution = {Technical University of Denmark (DTU)},
  year        = {2013},
  type        = {Reference Wind Turbine Report},
  url         = {https://orbit.dtu.dk/en/publications/the-dtu-10-mw-reference-wind-turbine}
}

@techreport{Jonkman_et_al_2009,
  title       = {Definition of a 5-MW Reference Wind Turbine for Offshore System Development},
  author      = {Jonkman, J. and Butterfield, S. and Musial, W. and Scott, G.},
  institution = {National Renewable Energy Laboratory (NREL)},
  type        = {Technical Report NREL/TP-500-38060},
  year        = {2009},
  month       = feb,
  note        = {Prepared under Task No. WER5.3301},
  url         = {https://www.nrel.gov/docs/fy09osti/38060.pdf}
}

@inproceedings{Meyers_Meneveau_2010_AIAA,
  title     = {Large eddy simulation of large wind-turbine arrays in the atmospheric boundary layer},
  author    = {Meyers, Johan and Meneveau, Charles},
  booktitle = {Proceedings of the AIAA Conference},
  year      = {2010},
  publisher = {American Institute of Aeronautics and Astronautics}
}

@article{Ayala_et_al_2024_MOSD,
  title   = {A Moving Surface Drag Model for LES of Wind Over Waves},
  author  = {Ayala, Manuel and Sadek, Zein and Fer\v{c}\'{a}k, Ond\v{r}ej and Cal, Ra\'{u}l Bayo\'{a}n and Gayme, Dennice F. and Meneveau, Charles},
  journal = {Boundary-Layer Meteorology},
  year    = {2024},
  doi     = {10.1007/s10546-024-00886-1}
}

@article{Moeng_1984,
  title   = {A large-eddy-simulation model for the study of planetary boundary-layer turbulence},
  author  = {Moeng, Chin-Hoh},
  journal = {Journal of the Atmospheric Sciences},
  volume  = {41},
  number  = {13},
  pages   = {2052--2062},
  year    = {1984},
  doi     = {10.1175/1520-0469(1984)041<2052:ALESMF>2.0.CO;2}
}

@book{Tennekes_Lumley_1972,
  title     = {A First Course in Turbulence},
  author    = {Tennekes, Henk and Lumley, John L.},
  publisher = {MIT Press},
  address   = {Cambridge, MA},
  year      = {1972}
}

@article{BouZeid_Meneveau_Parlange_2005,
  title   = {A scale-dependent Lagrangian dynamic model for large eddy simulation of complex turbulent flows},
  author  = {Bou-Zeid, Elie and Meneveau, Charles and Parlange, Marc},
  journal = {Physics of Fluids},
  volume  = {17},
  number  = {2},
  pages   = {025105},
  year    = {2005},
  doi     = {10.1063/1.1839152}
}

@techreport{Allen_et_al_UMaine_VolturnUS_15MW,
  title       = {Definition of the {UMaine} {VolturnUS-S} Reference Platform Developed for the {IEA} Wind 15-Megawatt Offshore Reference Wind Turbine},
  author      = {Allen, Christopher and Viselli, Anthony and Dagher, Habib and Goupee, Andrew and Gaertner, Evan and Abbas, Nikhar and Hall, Matthew and Barter, Garrett},
  institution = {National Renewable Energy Laboratory},
  address     = {Golden, CO},
  number      = {NREL/TP-5000-76773},
  year        = {2020}
}

@article{Porchetta2019,
  author    = {S. Porchetta and O. Temel and D. Mu{\~n}oz-Esparza and J. Reuder and J. Monbaliu and J. van Beeck and others},
  title     = {A New Roughness Length Parameterization Accounting for Wind--Wave (Mis)alignment},
  journal   = {Atmospheric Chemistry and Physics},
  volume    = {19},
  pages     = {6681--6700},
  year      = {2019},
  doi       = {10.5194/acp-19-6681-2019}
}

@article{Yang2014,
  author    = {D. Yang and C. Meneveau and L. Shen},
  title     = {Effect of downwind swells on offshore wind energy harvesting -- a large-eddy simulation study},
  journal   = {Renewable Energy},
  volume    = {70},
  pages     = {11--23},
  year      = {2014}
}

@article{AlSam2015,
  author    = {A. AlSam and R. Szasz and J. Revstedt},
  title     = {The Influence of Sea Waves on Offshore Wind Turbine Aerodynamics},
  journal   = {Journal of Energy Resources Technology},
  volume    = {137},
  number    = {5},
  pages     = {051209},
  year      = {2015}
}

@article{Subbulakshmi2022,
  author    = {A. Subbulakshmi and M. Verma and M. Keerthana and S. Sasmal and P. Harikrishna and S. Kapuria},
  title     = {Recent Advances in Experimental and Numerical Methods for Dynamic Analysis of Floating Offshore Wind Turbines -- An Integrated Review},
  journal   = {Renewable and Sustainable Energy Reviews},
  volume    = {164},
  pages     = {112525},
  year      = {2022}
}

@article{Sant2015,
  author    = {T. Sant and D. Bonnici and R. Farrugia and D. Micallef},
  title     = {Measurements and Modelling of the Power Performance of a Model Floating Wind Turbine under Controlled Conditions},
  journal   = {Wind Energy},
  volume    = {18},
  number    = {5},
  pages     = {811--834},
  year      = {2015}
}

@article{Wen2017,
  author    = {B. Wen and X. Tian and X. Dong and Z. Peng and W. Zhang},
  title     = {Influences of Surge Motion on the Power and Thrust Characteristics of an Offshore Floating Wind Turbine},
  journal   = {Energy},
  volume    = {141},
  pages     = {2054--2068},
  year      = {2017}
}

@article{Wen2018,
  author    = {B. Wen and X. Dong and X. Tian and Z. Peng and W. Zhang and K. Wei},
  title     = {The Power Performance of an Offshore Floating Wind Turbine in Platform Pitching Motion},
  journal   = {Energy},
  volume    = {154},
  pages     = {508--521},
  year      = {2018}
}

@article{Rockel2014,
  author    = {S. Rockel and E. Camp and J. Schmidt and J. Peinke and R. B. Cal and M. H{\"o}lling},
  title     = {Experimental Study on Influence of Pitch Motion on the Wake of a Floating Wind Turbine Model},
  journal   = {Energies},
  volume    = {7},
  number    = {4},
  pages     = {1954--1985},
  year      = {2014}
}

@article{energies2017_10_908,
  author  = {Bastankhah, Majid and Port{\'e}-Agel, Fernando},
  title   = {Energies},
  journal = {Energies},
  year    = {2017},
  volume  = {10},
  number  = {7},
  pages   = {908},
  doi     = {10.3390/en10070908},
  publisher = {MDPI},
  url     = {https://www.mdpi.com}
}

@article{wei2022phase,
  author  = {Wei, Nathaniel J. and Dabiri, John O.},
  title   = {Phase-averaged dynamics of a periodically surging wind turbine},
  journal = {Journal of Renewable and Sustainable Energy},
  year    = {2022},
  volume  = {14},
  pages   = {013305},
  doi     = {10.1063/5.0076029},
  url     = {https://doi.org/10.1063/5.0076029}
}

@article{messmer2024enhanced,
  author  = {Messmer, Thomas and H{\"o}lling, Michael and Peinke, Joachim},
  title   = {Enhanced recovery caused by nonlinear dynamics in the wake of a floating offshore wind turbine},
  journal = {Journal of Fluid Mechanics},
  year    = {2024},
  volume  = {984},
  pages   = {A66},
  doi     = {10.1017/jfm.2024.175},
  url     = {https://doi.org/10.1017/jfm.2024.175}
}

@article{duan2025floating,
  author  = {Duan, Guiyue and Gattari, Daniele and Port{\'e}-Agel, Fernando},
  title   = {Theoretical and experimental study on power performance and wake characteristics of a floating wind turbine under pitch motion},
  journal = {Applied Energy},
  year    = {2025},
  volume  = {378},
  pages   = {124767},
  doi     = {10.1016/j.apenergy.2024.124767},
  url     = {https://doi.org/10.1016/j.apenergy.2024.124767}
}

@article{li2024resolvent,
  author  = {Li, Zhaobin and Yang, Xiaolei},
  title   = {Resolvent-based motion-to-wake modelling of wind turbine wakes under dynamic rotor motion},
  journal = {Journal of Fluid Mechanics},
  year    = {2024},
  volume  = {980},
  pages   = {A48},
  doi     = {10.1017/jfm.2023.1097},
  url     = {https://doi.org/10.1017/jfm.2023.1097}
}

@misc{IEA15MW_ROSCO,
  author       = {Barter, Garrett and Bortolotti, Pietro and Gaertner, Evan and others},
  title        = {IEA-15-240-RWT: Reference Wind Turbine Repository},
  year         = {2025},
  doi          = {10.5281/zenodo.17449604},
  url          = {https://doi.org/10.5281/zenodo.17449604}
}

@article{zhou2025optimized,
  title={Optimized Structures for Passive Vibration Control of Floating Vertical-Axis Wind Turbine},
  author={Zhou, Fang and Yao, Zhiyuan and Yang, Mengyao and Chen, Jinbo and Zhao, Rui and Zhang, Yongfei},
  journal={Energies},
  volume={18},
  number={24},
  pages={6560},
  year={2025},
  doi={10.3390/en18246560}
}

@article{Vire2026FOWE,
  author       = {Axelle Vir{\'e} and Delphine De Tavernier and Dominic von Terzi},
  title        = {Floating Offshore Wind Energy: Challenges and Research Needs in Fluid Mechanics},
  journal      = {Flow, Turbulence and Combustion},
  year         = {2026},
  volume       = {116},
  number       = {2},
  pages        = {30},
  doi          = {10.1007/s10494-025-00717-9},
  url          = {https://doi.org/10.1007/s10494-025-00717-9}
}

@article{Dabiri2020Betz,
  author  = {John O. Dabiri},
  title   = {Theoretical framework to surpass the Betz limit using unsteady fluid mechanics},
  journal = {Physical Review Fluids},
  volume  = {5},
  number  = {2},
  pages   = {022501(R)},
  year    = {2020},
  month   = {February},
  doi     = {10.1103/PhysRevFluids.5.022501}
}

@article{Chattot2014ActuatorDisk,
  author  = {Jean-Jacques Chattot},
  title   = {Actuator Disk Theory—Steady and Unsteady Models},
  journal = {Journal of Solar Energy Engineering},
  volume  = {136},
  number  = {3},
  year    = {2014},
  doi     = {10.1115/1.4026947}
}

@article{Ayala2026waveRoughness,
  author  = {Manuel Ayala and Dennice F. Gayme and Charles Meneveau},
  title   = {Surface Wave-Aerodynamic Roughness Length Model for Air--Sea Interactions},
  journal = {Journal of the Atmospheric Sciences},
  pages   = {281--299},
  year    = {2026},
  doi     = {10.1175/JAS-D-25-0146.1},
  url     = {https://doi.org/10.1175/JAS-D-25-0146.1}
}

@article{Charnock1955,
  author  = {H. Charnock},
  title   = {Wind stress on a water surface},
  journal = {Quarterly Journal of the Royal Meteorological Society},
  volume  = {81},
  pages   = {639--640},
  year    = {1955},
  doi     = {10.1002/qj.49708135027}
}

@article{MoninObukhov1954,
  author  = {Monin, A. S. and Obukhov, A. M.},
  title   = {Basic laws of turbulent mixing in the surface layer of the atmosphere},
  journal = {Trudy Geofizicheskogo Instituta AN SSSR},
  year    = {1954},
  volume  = {24},
  pages   = {163--187}
}

@article{Wu2011,
  author  = {Wu, Yu-Ting and Port{\'e}-Agel, Fernando},
  title   = {Large-Eddy Simulation of Wind-Turbine Wakes: Evaluation of Turbine Parametrisations},
  journal = {Boundary-Layer Meteorology},
  year    = {2011},
  volume  = {138},
  number  = {3},
  pages   = {345--366},
  doi     = {10.1007/s10546-010-9569-x},
  publisher = {Springer}
}

@article{Tran2026,
  author = {Tran, Dung and Sarker, Doyal and Mohsin, Kazi and Rahman, Saad and Sakif, Md and Ngo, Tri and Bergua, Roger and Zalkind, Daniel and Jonkman, Jason and Das, Tuhin},
  title = {Modeling, Validation, and Control of the IEA-15 MW Reference Wind Turbine and VolturnUS-S Platform},
  journal = {Wind Energy},
  year = {2026},
  volume = {29},
  number = {5},
  pages = {e70115},
  doi = {10.1002/we.70115},
  url = {https://doi.org/10.1002/we.70115}
}

@article{Howland2016Wake,
  author = {Howland, Michael F. and Bossuyt, Juliaan and Martínez-Tossas, Luis A. and Meyers, Johan and Meneveau, Charles},
  title = {Wake structure in actuator disk models of wind turbines in yaw under uniform inflow conditions},
  journal = {Journal of Renewable and Sustainable Energy},
  volume = {8},
  number = {4},
  pages = {043301},
  year = {2016},
  publisher = {AIP Publishing},
  doi = {10.1063/1.4955307}
}

@article{espana2011spatial,
  author    = {G. Espa{\~n}a and S. Aubrun and S. Loyer and P. Devinant},
  title     = {Spatial study of the wake meandering using modelled wind turbines in a wind tunnel},
  journal   = {Wind Energy},
  volume    = {14},
  number    = {2},
  pages     = {239--251},
  year      = {2011},
  publisher = {Wiley Online Library}
}

@manual{Ansys2025,
  author       = {{Ansys}},
  title        = {Ansys Fluent User's Guide (Release 2025 R2)},
  organization = {Ansys, Inc.},
  address      = {Canonsburg, PA},
  year         = {2025},
  note         = {Available at: \url{https://ansys.com} (accessed May 30, 2026)}
}
\end{document}